\PassOptionsToPackage{unicode}{hyperref}
\PassOptionsToPackage{hyphens}{url}
\PassOptionsToPackage{dvipsnames,svgnames,x11names}{xcolor}
\documentclass[
  12pt]{article}

\usepackage{geometry} 
\usepackage{rotating}
\usepackage[export]{adjustbox}
\usepackage {mathpazo}
\usepackage{tabularx}
\usepackage{placeins}
\usepackage{booktabs}
\usepackage{amsthm}
\usepackage{float}
\usepackage{setspace} 
\usepackage{indentfirst}
\usepackage[skip=5pt, indent=15pt]{parskip}
\usepackage{graphicx} 
\usepackage{authblk} 
\usepackage{hyperref}
\usepackage{yhmath}
\usepackage{makecell}
\hypersetup{colorlinks,linkcolor={blue},citecolor={blue},urlcolor={black}}
\usepackage[strict]{changepage} 

\usepackage{amsthm}
\usepackage{bbm}
\usepackage{nicefrac}
\usepackage{enumitem}
\usepackage{amsfonts}
\usepackage{breakcites}
\usepackage{mathtools}
\usepackage{relsize}
\usepackage{subfig}
\usepackage{makecell}
\usepackage{pdflscape}
\usepackage{listings}
\usepackage{lipsum}
\usepackage{bm}

\usepackage{scalerel,stackengine}
\stackMath
\newcommand\reallywidehat[1]{%
\savestack{\tmpbox}{\stretchto{%
  \scaleto{%
    \scalerel*[\widthof{\ensuremath{#1}}]{\kern-.6pt\bigwedge\kern-.6pt}%
    {\rule[-\textheight/2]{1ex}{\textheight}}
  }{\textheight}%
}{0.5ex}}%
\stackon[1pt]{#1}{\tmpbox}%
}
\newtheorem{proposition}{Proposition}
\newtheorem{remark}{Remark}
\usepackage[ruled,vlined, linesnumbered]{algorithm2e}

\DeclareMathOperator*{\argmax}{\arg\!\max}
\DeclareMathOperator*{\argmin}{\arg\!\min}

\usepackage{amsmath,amssymb}
\usepackage{iftex}
\ifPDFTeX
  \usepackage[T1]{fontenc}
  \usepackage[utf8]{inputenc}
  \usepackage{textcomp} 
\else 
  \usepackage{unicode-math}
  \defaultfontfeatures{Scale=MatchLowercase}
  \defaultfontfeatures[\rmfamily]{Ligatures=TeX,Scale=1}
\fi
\usepackage{lmodern}
\ifPDFTeX\else  
\fi
\IfFileExists{upquote.sty}{\usepackage{upquote}}{}
\IfFileExists{microtype.sty}{
  \usepackage[]{microtype}
  \UseMicrotypeSet[protrusion]{basicmath} 
}{}
\makeatletter
\@ifundefined{KOMAClassName}{
  \IfFileExists{parskip.sty}{%
    \usepackage{parskip}
  }{
    \setlength{\parindent}{0pt}
    \setlength{\parskip}{6pt plus 2pt minus 1pt}}
}{
  \KOMAoptions{parskip=half}}
\makeatother
\usepackage{xcolor}
\makeatletter
\ifx\paragraph\undefined\else
  \let\oldparagraph\paragraph
  \renewcommand{\paragraph}{
    \@ifstar
      \xxxParagraphStar
      \xxxParagraphNoStar
  }
  \newcommand{\xxxParagraphStar}[1]{\oldparagraph*{#1}\mbox{}}
  \newcommand{\xxxParagraphNoStar}[1]{\oldparagraph{#1}\mbox{}}
\fi
\ifx\subparagraph\undefined\else
  \let\oldsubparagraph\subparagraph
  \renewcommand{\subparagraph}{
    \@ifstar
      \xxxSubParagraphStar
      \xxxSubParagraphNoStar
  }
  \newcommand{\xxxSubParagraphStar}[1]{\oldsubparagraph*{#1}\mbox{}}
  \newcommand{\xxxSubParagraphNoStar}[1]{\oldsubparagraph{#1}\mbox{}}
\fi
\makeatother

\usepackage{longtable,booktabs,array}
\usepackage{calc} 
\usepackage{etoolbox}
\makeatletter
\patchcmd\longtable{\par}{\if@noskipsec\mbox{}\fi\par}{}{}
\makeatother
\IfFileExists{footnotehyper.sty}{\usepackage{footnotehyper}}{\usepackage{footnote}}
\makesavenoteenv{longtable}
\usepackage{graphicx}
\makeatletter
\def\maxwidth{\ifdim\Gin@nat@width>\linewidth\linewidth\else\Gin@nat@width\fi}
\def\maxheight{\ifdim\Gin@nat@height>\textheight\textheight\else\Gin@nat@height\fi}
\makeatother
\setkeys{Gin}{width=\maxwidth,height=\maxheight,keepaspectratio}
\makeatletter
\def\fps@figure{htbp}
\makeatother

\makeatletter
\@ifpackageloaded{caption}{}{\usepackage{caption}}
\AtBeginDocument{%
\ifdefined\contentsname
  \renewcommand*\contentsname{Table of contents}
\else
  \newcommand\contentsname{Table of contents}
\fi
\ifdefined\listfigurename
  \renewcommand*\listfigurename{List of Figures}
\else
  \newcommand\listfigurename{List of Figures}
\fi
\ifdefined\listtablename
  \renewcommand*\listtablename{List of Tables}
\else
  \newcommand\listtablename{List of Tables}
\fi
\ifdefined\figurename
  \renewcommand*\figurename{Figure}
\else
  \newcommand\figurename{Figure}
\fi
\ifdefined\tablename
  \renewcommand*\tablename{Table}
\else
  \newcommand\tablename{Table}
\fi
}
\@ifpackageloaded{float}{}{\usepackage{float}}
\floatstyle{ruled}
\@ifundefined{c@chapter}{\newfloat{codelisting}{h}{lop}}{\newfloat{codelisting}{h}{lop}[chapter]}
\floatname{codelisting}{Listing}

\makeatother
\makeatletter
\@ifpackageloaded{caption}{}{\usepackage{caption}}
\@ifpackageloaded{subcaption}{}{\usepackage{subcaption}}
\makeatother

\ifLuaTeX
  \usepackage{selnolig}  
\fi
\usepackage[round,authoryear]{natbib}
\usepackage{bookmark}

\IfFileExists{xurl.sty}{\usepackage{xurl}}{} 
\hypersetup{
  pdftitle={Title},
  pdfauthor={Author 1; Author 2},
  pdfkeywords={3 to 6 keywords, that do not appear in the title},
  colorlinks=true,
  linkcolor={blue},
  filecolor={Maroon},
  citecolor={Blue},
  urlcolor={Blue},
  pdfcreator={LaTeX via pandoc}}

\newcommand{\anon}{1}

\begin{document}

\def\spacingset#1{\renewcommand{\baselinestretch}%
{#1}\small\normalsize} \spacingset{1}


\if1\anon
{
  \title{\bf Ensemble Control Variates}
  \author{Minh-Long Nguyen, Christopher Drovandi, Leah F. South\hspace{.2cm}\\
    School of Mathematical Sciences, Queensland University of Technology, Australia \\
    \textit{Corresponding email: \href{mailto:minhlong.nguyen@hdr.qut.edu.au}{minhlong.nguyen@hdr.qut.edu.au}}}
  \maketitle
} \fi

\if0\anon
{
  \bigskip
  \bigskip
  \bigskip
  \begin{center}
    {\LARGE\bf Title}
\end{center}
  \medskip
} \fi

\bigskip
\begin{abstract}
Control variates have become an increasingly popular variance‑reduction technique in Bayesian inference. Many broadly applicable control variates are based on the Langevin–Stein operator, which leverages gradient information available from gradient‑based samplers to produce variance‑reduced estimators of expectations. These control variates typically require optimising over a function $u(\theta)$ within a user‑defined functional class $\mathcal{G}$, such as the space of $Q$th‑order polynomials or a reproducing kernel Hilbert space. We propose using averaging‑based ensemble learning to construct Stein‑based control variates. While the proposed framework is broadly applicable, we focus on ensembles constructed from zero‑variance control variates (ZVCV), a popular parametric approach based on solving a linear approximation problem that can easily be over‑parameterised in medium-to-high dimensional settings. A common remedy is to use regularised ZVCV via penalised regression, but these methods can be prohibitively slow. We introduce ensemble ZVCV methods based on ensembles of OLS estimators and evaluate the proposed methods against established methods in the literature in a simulation study. Our results show that ensemble ZVCV methods are more statistically efficient than regularised ZVCV methods, while also being substantially faster. This work opens a new direction for constructing broadly applicable control variate techniques via ensemble learning.

\end{abstract}

\noindent%
\vfill

\section{Introduction}

Given a random sample $\{\theta_1, \ldots, \theta_S\} \subseteq \Theta \subseteq \mathbb{R}^d$ from a continuous probability distribution $\pi(\theta)$ and a square-integrable $f(\theta)$, our work focuses on improving the statistical efficiency of the vanilla Monte Carlo estimator $\hat{I}_{\text{MC}}$:
\begin{equation}
    \hat{I}_{\text{MC}} \overset{\Delta}{=} \frac{\sum_{i=1}^Sf(\theta_i)}{S}.
\end{equation}

This estimator generally has the standard convergence rate of $\mathcal{O}(S^{-1/2})$ \citep{leluc2023speeding}, implying that to reduce the root-mean-squared error (RMSE) of the vanilla estimator by a factor of $k$, the sample size must be increased by $\mathcal{O}(k^2)$. This holds true in independent and identically distributed (i.i.d.) sampling or popular dependent sampling frameworks such as Markov chain Monte Carlo (MCMC, \citealt{brooks2011handbook}). However, in many applications, particularly in Bayesian inference, sampling from $\pi(\theta)$ can be prohibitively expensive. Therefore, it is highly desirable to improve the statistical efficiency of the vanilla estimator for any fixed sample sizes.

Various so-called variance reduction methods have been proposed in the literature. We focus on an increasingly popular variance-reduction technique known as control variates; see \citet[Ch.~8]{mcbook} for an introduction. This technique involves subtracting a synthesised square-integrable $\Tilde{f}(\theta)$ with a known mean $\alpha$, which we refer to as a control variate, from ${f}(\theta)$, resulting in a new estimator $\hat{I}_{\text{CV}}$:
\begin{align}
\label{def:cv}
    \hat{I}_{\text{CV}} &\overset{\Delta}{=} \frac{\sum_{i=1}^S\big(f(\theta_i) - \Tilde{f}(\theta_i)\big) + \alpha}{S} \nonumber \\
    &\text{ s.t. } \mathbb{E}_{\pi}[\Tilde{f}(\theta)] = \alpha,
\end{align}
that has the same expectation but a smaller variance if $\mathbb{V}_{\pi}[f(\theta) - \Tilde{f}(\theta)] < \mathbb{V}_{\pi}[f(\theta)]$.

Constructing a control variate estimator used to be problem-specific \citep{hammersley1964general, ripley1987stochastic}. Recent work has led to the development of broadly applicable gradient-based control variates based on the Langevin-Stein operator $\mathcal{L}$ \citep{stein1972bound}, which we refer to as Stein-based control variates; see \citealt{south2022postprocessing} and \citealt{south2024control} for reviews. These types of control variates utilise the gradient of the log-target density, which is available from popular gradient-based samplers such as the Metropolis-adjusted Langevin algorithm (MALA, \citealt{rossky1978brownian}) and the no-u-turn sampler (NUTS, \citealt{hoffman2014no, carpenter2017stan}). These have been applied in various settings, such as posterior mean estimation, model evidence estimation \citep{oates2017control, south2023regularized}, variational inference \citep{ng2024pathwise}, to name just a few.

Stein-based control variates involve optimising over the choice of function $u(\theta) \in \mathcal{G}$, and there are various options for $\mathcal{G}$. Zero-variance control variates (ZVCV, \citealt{assaraf1999zero, mira2013zero}) select $u(\theta)$ from the class of $Q$th order polynomial functions in $\theta$. Control functionals (CF, \citealt{oates2017control}) select $u(\theta)$ from a reproducing kernel Hilbert space (RKHS) and have been extended to vector-valued and multi-level settings \citep{sun2023vector,  li2023multilevel}. \citealt{south2022semi} propose semi-exact control functionals (SECF), where the Gaussian process prior in CF is partially informed by a parametric model, particularly that from ZVCV. The function $u(\theta)$ can be parameterised by a neural network \citep{wan2019neural, sun2023meta}. Weight optimisation or hyper-parameter tuning are generally required for these methods, and in the context of ZVCV this is typically performed via linear regression. Outside the Langevin-Stein framework, control variates have been constructed for reversible MCMC samplers \citep{dellaportas2012control}.

We propose using ensemble learning for constructing control variates. This technique consists of combining multiple weak learners to construct a stronger learner and is commonly used in many machine learning applications \citep{mienye2022survey}. An example of ensemble learning in control variates is SECF, which can be viewed as a boosting-style (i.e., iterative improvement) ensemble control variate, where the non-parametric CF is partially informed by a parametric ZVCV model. In this work, we instead explore averaging-based ensemble learning for Stein-based control variates, which, to the best of our knowledge, has not yet been applied in this context. Our use of the term ``ensemble'' differs from that of \citet{geffner2018using} and \citet{pham2022ensemble}, who define an ensemble control variate as a weighted average of pre-defined control variates. When the weights are estimated jointly, such an estimator is a single learner; in that sense, ZVCV, which linearly combines a pre-defined set of polynomial control variates, is already an ensemble. By contrast, our ensemble control variates average several learners, each fitted to a different subset of the candidate control variates.

While our proposed framework is general, we primarily focus on ZVCV in this work. Performing ZVCV is equivalent to solving a multiple linear regression problem, for which various criteria can be used. A common choice is the ordinary least squares (OLS) criterion. However, as the number of polynomial coefficients $J$ grows factorially in $Q$ and $d$, standard ZVCV via OLS becomes non-identifiable in many applications where $S < J$. This can still be solved in a ridgeless fashion using the Moore--Penrose pseudo-inverse; however, it may result in a variance-inflated control variate estimator \citep{leluc2021control}, which is undesirable. To address this, \cite{south2023regularized, leluc2021control} propose regularised ZVCV methods that apply penalised regression techniques such as LASSO \citep{tibshirani1996regression} and ridge regression \citep{hoerl1970ridge} to the linear approximation task. However, these methods can be prohibitively expensive.

Ensemble OLS estimators can address over-parameterised linear approximation problems. Restricting each OLS learner to a subset of the covariates is itself a form of regularisation, as the coefficients of the unselected covariates are constrained to zero, while averaging over learners fitted to different subsets allows every covariate to contribute to the final estimator. This construction is conceptually related to random forests \citep{breiman2001random} and to bagged OLS with feature randomness \citep{lejeune2020implicit, patil2023generalized, patil2024asymptotically}, although we subsample only the covariates, not the samples, so that every learner uses the full sample. Under standard regression assumptions and asymptotic sub-sampling conditions, a large ensemble of bagged OLS predictors with feature randomness may exhibit an implicit regularisation equivalent to that of the optimally tuned ridge regression \citep{lejeune2020implicit}. In other words, one may achieve the statistical benefits of ridge regression without explicitly performing it. Although the required assumptions are typically violated in control-variates contexts (and in ZVCV), this motivates our proposal of ensemble ZVCV estimators. The proposed ensemble ZVCV methods are expected to perform at least comparably to regularised ZVCV techniques in terms of statistical efficiency, while being substantially faster, particularly when $J$ is large relative to $S$.


This manuscript is organised as follows. Section \ref{section:bg} provides the necessary background. Section \ref{section:ecv} introduces the proposed approaches, characterises the variance of the resulting estimators and presents heuristics for selecting hyper-parameters. Section \ref{section:sstudy} compares the proposed methods and established Stein-based control variate techniques in a simulation study. Finally, a conclusion is given in Section \ref{section:conclusion} with some directions for future work.

\section{Background}
\label{section:bg}

This section provides necessary background on the Langevin-Stein operator and ZVCV.

We assume that a sample $\{\theta_1,...,\theta_S\}$ is available as per earlier. For ZVCV, we assume the tails of $\pi(\theta)$ decay faster than polynomially \citep{south2024control}. Let $\{\theta_{[1]},\ldots,\theta_{[d]}\}$ denote the $d$ dimensions of $\theta$. For simplicity, we assume both $f(\theta)$ and $\Tilde{f}(\theta)$ are scalar-valued. Let $\textbf{f}_S, \Tilde{\textbf{f}}_S \in \mathbb{R}^S$ be the vectorised evaluations of square-integrable $f(\theta)$ and $\Tilde{f}(\theta)$. Let $\textbf{g}_S$ be the vectorised evaluation of $\nabla_{\theta}\log\pi(\theta)$. Given an arbitrary vector \textbf{v}, we use $\textbf{v}[i]$ to denote the $i$-th entry of $\textbf{v}$. Given a matrix $\textbf{Z}$, we use $\textbf{Z}[i,j]$ to denote the entry in the $i$th-row and $j$th-column, $\textbf{Z}[i,]$ to denote the $i$th row, and $\textbf{Z}[,j]$ to denote the $j$th column.

\subsection{The Langevin-Stein Operator}
\label{section:stein}
Many recently proposed control variates techniques are based on the Langevin-Stein operator $\mathcal{L}$,  which can be used to construct zero-mean control variates for a class of function $\mathcal{G}$ \citep{oates2017control}:
\begin{equation}
    \label{eqn:steinoperator}
    \mathbb{E}_{\pi}[\mathcal{L}u(\theta)] = 0; \; \forall u(\theta) \in \mathcal{G}, \; \theta \sim \pi(\theta).
\end{equation}
A popular option is the first-order Langevin-Stein operator $\mathcal{L}_1$, which CF is based on:
\begin{equation}
    \mathcal{L}_1 u(\theta) = \nabla_{\theta}\cdot u(\theta) + u(\theta) \cdot \nabla_{\theta}\log \pi(\theta),
\end{equation}
where $u(\theta)$ is a vector-valued function $u: \mathbb{R}^d \rightarrow \mathbb{R}^d$ \citep{oates2017control}. A second popular option is the second-order Langevin-Stein operator $\mathcal{L}_2$, which ZVCV and regularised ZVCV are based on:
\begin{equation}
    \mathcal{L}_2 u(\theta) = \Delta_{\theta}u(\theta) + \nabla_{\theta}u(\theta) \cdot \nabla_{\theta}\log \pi(\theta),
\end{equation}
where $u(\theta)$ is a real-valued function, and $\Delta_{\theta}$ denotes the sum of second-order partial derivatives with respect to (w.r.t.) each $\theta$.

Both operators satisfy Equation \eqref{eqn:steinoperator} under standard boundary conditions \citep{mira2013zero, Oates2019, south2022semi}. Since $\mathcal{L}_1 u(\theta) = \pi(\theta)^{-1}\nabla_{\theta}\cdot\big(\pi(\theta)u(\theta)\big)$, $\mathbb{E}_{\pi}[\mathcal{L}_1u(\theta)]$ is the integral of a divergence, which vanishes when $\pi(\theta)u(\theta)$ decays sufficiently quickly in the tails. The operator $\mathcal{L}_2$ is obtained by applying $\mathcal{L}_1$ to the gradient field $\nabla_\theta u(\theta)$, and is the generator of the overdamped Langevin diffusion with invariant measure $\pi$ \citep{roberts1996exponential}, which is why $\mathcal{L}_2u(\theta)$ has zero mean under $\pi$.

In this work, we consider control variates $\tilde{f}(\theta)$ of the form $\mathcal{L}_2u(\theta)+\alpha$.

\subsection{Zero-Variance Control Variates}
\label{section:zvcv}

In ZVCV \citep{assaraf1999zero, mira2013zero}, $u(\theta)$ is selected from the class of $Q$th order polynomials in $\theta$ ($\mathcal{P}^Q$). In particular,
\begin{equation}
    u(\theta) = \sum_{\bm{\alpha^j} \in \mathbb{N}_0^d, 0 < |\bm{\alpha^j}| \leq Q} \beta_{\bm{\alpha^j}}\theta^{\bm{\alpha^j}},
\end{equation}
where $\bm{\alpha^j} = \{\bm\alpha^j_1, \ldots,\bm\alpha^j_d\} \subset \mathbb{N}_0^d$ such that $0 < |\bm{\alpha^j}| \triangleq \sum_{i=1:d} \bm\alpha^j_i \leq Q$, $j \in \{1,\cdots,J\}$, $J = {Q+d\choose d}-1$; $\theta^{\bm{\alpha^j}} \triangleq \prod_{i=1:d}\theta_{[i]}^{\bm\alpha^j_i}$ is written in multi-index notation; and $\beta_{\bm{\alpha^j}} \in \mathcal{\mathbb{R}}$ is the coefficient corresponding to the monomial $\theta^{\bm{\alpha^j}}$. Let \textbf{Z} denote an $S\times J$ matrix, which we refer to as the ZVCV design matrix, where

\begin{equation}
\textbf{Z}[i,] = \Bigg\{ \mathcal{L}_2 \theta^{\alpha^1}, \; \mathcal{L}_2 \theta^{\alpha^2}, \; \dots, \; \mathcal{L}_2 \theta^{\alpha^J} \Bigg\}^\top \Bigg|_{\theta = \theta_i}; \; i \in \{1,...,S\} \;\; \triangleq \;\; \phi(\theta) \Bigg|_{\theta = \theta_i}.
\end{equation}

Performing ZVCV is equivalent to solving the linear approximation problem $\textbf{f}_S \approx \textbf{Z}\bm\beta+\alpha$, $\alpha \in \mathbb{R}$. If $S > J$, $\bm\beta$ can be optimised via the OLS criterion:

\begin{equation}
    \label{eqn:ols}
    \{\hat{\alpha}_{\text{OLS}}, \hat{\bm\beta}_{\text{OLS}}\} = \argmin_{\alpha \in \mathbb{R}, \bm\beta \in \mathbb{R}^J} \big\|\textbf{Z}\bm\beta + \alpha - \textbf{f}_S\big\|_2^2.
\end{equation}

This results in the ZVCV estimator $\hat{I}_{\text{ZV}}$:
\begin{align}
\label{eqn:zvcv}
\hat{I}_{\text{ZV}} &=\frac{1}{S}\sum_{i=1}^S \big(f(\theta_i)  - \phi^\top(\theta_i)\hat{\bm\beta}_{\text{OLS}}-\hat{\alpha}_{\text{OLS}} + \hat{\alpha}_{\text{OLS}}\big),
\\
&= \frac{\mathbbm{1}_{S \times 1}^\top\big(\textbf{f}_S - \textbf{Z}\hat{\bm\beta}_{\text{OLS}}\big)}{S}\\
\label{eqn:intercept}
&= \hat{\alpha}_{\text{OLS}} \; \; \text{{(as the sum of regression residuals is 0)}},
\end{align}

where $\mathbbm{1}_{S \times 1}$ denotes a vector of ones of size $S$. Here we use the same samples in both the OLS estimation and the control variate estimation stages. This may introduce some bias, which is generally negligible for practical purposes.  However, we still have consistency \citep{mira2013zero, portier2019monte}.

The pseudocode for the ZVCV algorithm is given in Algorithm \ref{alg:zvcv}. The QR decomposition can be used to solve the OLS problem, offering improved numerical stability compared to explicitly computing a matrix inverse. As the QR factorisation is performed only once, control variate estimates for multiple expectations can be obtained simultaneously.

\begin{algorithm}[tb]
   \caption{The ZVCV Algorithm (fixed $Q$)}
   \label{alg:zvcv}
   \SetKwInput{KwData}{Input}
   \KwData{samples $\{\theta_1,...,\theta_S\}$, integrand evaluation
   $\textbf{f}_S$, gradient evaluation of log-target density $\textbf{g}_S$}

   Compute the ZVCV design matrix $\mathbf{Z}$ from $\{\theta_1,...,\theta_S\}$ and $\textbf{g}_S$\;
   Solve for $\{\hat{\alpha}_{\text{OLS}}, \hat{\bm\beta}_{\text{OLS}}\} = \argmin_{\alpha \in \mathbb{R}, \bm\beta \in \mathbb{R}^J} \big\|\textbf{Z}\bm\beta + \alpha - \textbf{f}_S\big\|_2^2$ \;
   Return $\hat{\alpha}_{\text{OLS}}$.
\end{algorithm}

An interesting property of ZVCV is that it can give a zero-variance estimator when $\pi(\theta)$ is Gaussian, and $f(\theta)$ is a $Q'$th-order polynomial function, $Q' \leq Q$ (hence the name ZVCV \citep{mira2013zero}). In other words, $\hat{I}_\text{ZV}$ can perfectly estimate $\mathbb{E}_{\pi}[f(\theta)]$ using a finite sample. This is particularly useful when $\pi(\theta)$ is approximately Gaussian and $f(\theta)$ is a low-order polynomial function (e.g., $f(\theta) = \theta$ as in posterior mean estimation).

For computational efficiency, ZVCV is often implemented with $Q = 1$ or $Q = 2$. Although increasing the polynomial order may improve statistical efficiency, it also increases $J$ factorially as $J = \binom{d+Q}{d} - 1$, potentially resulting in $S < J$ and rendering the OLS problem unidentifiable. Even when $S \approx J$, statistical performance can be poor. Consequently, ZVCV becomes challenging to apply for large dimension $d$ when a high polynomial order is required for good statistical performance. One potential remedy is to omit interaction terms in $u(\theta)$, reducing $J$ to $Qd$. However, this may incur a substantial loss of efficiency, particularly if the dimensions $\theta_{[1]}, \dots, \theta_{[d]}$ are correlated. This is because $\mathcal{L}_2$ couples the dimensions through $\nabla_\theta\log\pi(\theta)$. For example, if $\pi(\theta)$ is Gaussian with mean zero and precision matrix $\mathbf{P}$, then $\mathcal{L}_2\theta_{[1]}^2 = 2 - 2\theta_{[1]}\sum_{j=1}^d \mathbf{P}[1,j]\,\theta_{[j]}$, which contains the interaction terms $\theta_{[1]}\theta_{[j]}$ whenever $\mathbf{P}[1,j]\neq 0$. These terms cannot be cancelled without interaction monomials in $u(\theta)$, so ZVCV loses its zero-variance property for, e.g., $f(\theta) = \theta_{[1]}^2$.

A common remedy for this is regularised ZVCV techniques \citep{south2023regularized, leluc2021control}, based on penalised regression criteria such as LASSO and ridge regression. Particularly:

\begin{enumerate}
    \item LASSO-based regularised ZVCV:

    \begin{equation}
    \label{eqn:l1}
       \{\hat{\alpha}_{\text{LASSO}}, \hat{\bm\beta}_{\text{LASSO}}\} = \argmin_{\alpha\in \mathbb{R}, \beta \in \mathbb{R}^J} \big\|\textbf{Z}\bm\beta + \alpha - \textbf{f}_S\big\|_2^2 + \lambda\|\bm\beta\|_1.
    \end{equation}
    \item Ridge-regression-based regularised ZVCV:

    \begin{equation}
       \{\hat{\alpha}_{\text{Ridge}}, \hat{\bm\beta}_{\text{Ridge}}\} = \argmin_{\alpha\in \mathbb{R}, \beta \in \mathbb{R}^J} \big\|\textbf{Z}\bm\beta + \alpha - \textbf{f}_S\big\|_2^2 + \lambda\|\bm\beta\|_2^2.
    \end{equation}
\end{enumerate}

Both approaches require a non‑negative hyperparameter $\lambda$, which is typically selected via cross‑validation. However, cross‑validation can be computationally expensive when both $S$ and $J$ are moderately large. In some cases, LASSO may require many iterations to converge. Moreover, with multiple integrands, a separate regularised regression problem must be solved for each. As a result, simultaneous estimation of expectations cannot be performed on a single core.

\section{Ensemble Control Variates}
\label{section:ecv}

We introduce a novel Stein-based control variate framework based on averaging-based ensemble learning. While the proposed framework is broadly applicable, we mainly focus on ensembles constructed from ZVCV estimators.

We define a mapping $\mathbb{S}: \{1,...,J\} \rightarrow \{0,1\}^J$, where $\mathbb{S}[j] = 1$ implies the $j$th covariate is ``selected'', $j \in \{1,...,J\}$.

\subsection{Definition}
\label{section:def}

We consider an ensemble control variate $\Tilde{f}_{\mathrm{ens}}$ of the following form:
\begin{equation}
    \Tilde{f}_{\mathrm{ens}}(\theta) \triangleq \sum_{i=1}^k w_{i}\mathcal{L}_2u_i(\theta) + w_i\alpha_i, \label{eqn:def_ens}
\end{equation}

where $u_i(\theta)$ are selected from some classes of functions $\mathcal{G}_i$, and $\{w_1,\cdots,w_k\}$ are some weights. A simple option is to use normalised weights such that $\sum_{i=1}^k w_{i} = 1$, and $w_{i} \geq 0, i = 1,...,k$. Alternatively, $\{w_1,\cdots,w_k\}$ can be unbounded.

In this work, $u_i(\theta) \in \mathcal{P}^\mathcal{Q}$, where we refer to the proposed approaches as ensemble ZVCV methods. However, it is not strictly necessary for all $u_i(\theta)$ to belong to the same functional class. Candidates for $\mathcal{G}_i$ include, but are not limited to, $\mathcal{P}^\mathcal{Q}$, RKHS (e.g., the ensemble Nyström method \citep{kumar2009ensemble}), and neural networks. We leave this for future research.

To understand the role of the weights, consider fixed functions $u_1,\ldots,u_k$ and normalised weights, and let $R_i(\theta) \triangleq f(\theta) - \mathcal{L}_2u_i(\theta)$ with $\mathbb{V}_\pi[R_i(\theta)] < \infty$, $i = 1,\ldots,k$. The ensemble estimator of $\mathbb{E}_\pi[f(\theta)]$ is then $\sum_{i=1}^k w_i \frac{1}{S}\sum_{s=1}^S R_i(\theta_s)$, which is unbiased and, given an i.i.d.\ sample, has variance $\mathbf{w}^\top\bm\Sigma\mathbf{w}/S$, where $\mathbf{w} = (w_1,\ldots,w_k)^\top$ and $\bm\Sigma[i,j] = \mathrm{Cov}_\pi\big(R_i(\theta),R_j(\theta)\big)$. For MCMC samples, the same expression holds asymptotically, with $\bm\Sigma$ replaced by the matrix of asymptotic (long-run) covariances.

\begin{proposition}[Variance bounds]
\label{prop:error-bounds}
Let $V^\star=\min_{\mathbf{w}\in\mathcal{W}}\mathbf{w}^{\top}\bm{\Sigma}\mathbf{w}$, where $\mathcal{W}
= \left\{ w \in \mathbb{R}^k :\sum_{i=1}^k w_i = 1
\right\}$. If $\mathcal{W}$ contains the canonical basis vectors, then the variance of the optimally weighted ensemble estimator, $V^\star/S$, satisfies $V^\star/S\leq\frac{1}{S}\min_i\mathbb{V}_\pi[R_i(\theta)]$, and it is zero whenever some $R_i(\theta)$ has zero variance. For equal weights $w_i=1/k$, the variance of the ensemble estimator satisfies
\[
\frac{1}{Sk^2}\mathbbm{1}_{k\times1}^\top\bm\Sigma\mathbbm{1}_{k\times1}
\leq
\frac{1}{S}\max_i\mathbb{V}_\pi[R_i(\theta)].
\]
\end{proposition}
\begin{proof}
Since $\mathcal{W}$ contains each canonical basis vector $\mathbf{e}_i$ and $\mathbf{e}_i^\top\bm\Sigma\mathbf{e}_i=\mathbb{V}_\pi[R_i(\theta)]$, we have $V^\star\leq\min_i\mathbb{V}_\pi[R_i(\theta)]$. As $\bm\Sigma$ is positive semi-definite, $V^\star\geq0$, so $V^\star=0$ if $\mathbb{V}_\pi[R_i(\theta)]=0$ for some $i$. For equal weights, write $\sigma_i = \mathbb{V}_\pi[R_i(\theta)]^{1/2}$. By the Cauchy--Schwarz inequality, $\bm\Sigma[i,j] \leq \sigma_i\sigma_j$, so $\mathbbm{1}_{k\times1}^\top\bm\Sigma\mathbbm{1}_{k\times1}\leq\big(\sum_i\sigma_i\big)^2\leq k^2\max_i\sigma_i^2$. Dividing by $Sk^2$ gives the result.
\end{proof}

\begin{remark}
\label{rem:weights}
Proposition \ref{prop:error-bounds} treats the learners as fixed, whereas in practice both the learners and the weights are estimated from the same sample. It therefore describes the idealised behaviour of the two types of weights rather than giving a finite-sample guarantee: variance-optimised weights can do no worse than the best single learner, at the cost of estimating $\bm\Sigma$, while equal weights require no estimation and guard against the worst learner.
\end{remark}

\subsection{Ensemble Zero-Variance Control Variates}
\label{section:ezvcv}

For ensemble ZVCV approaches, $u_i(\theta) \in \mathcal{P}^\mathcal{Q}$, we can rewrite the control variate in Equation \ref{eqn:def_ens} as
\begin{align}
    \Tilde{f}_{\mathrm{ens}}(\theta) &= \sum_{i=1}^k w_{i}\phi^\top(\theta)\bm\beta_{i} + w_i\alpha_i\label{eqn:ens} \\
    &= \phi^\top(\theta)\sum_{i=1}^k w_{i}\bm\beta_{i}  + \sum_{i=1}^k w_i\alpha_i ,
\end{align}

where $\bm\beta_i$ is a vector of polynomial coefficients in $u_i(\theta)$, and $\alpha_i$ is the $i$th intercept. Here, $\Tilde{f}_{\mathrm{ens}}$ can be interpreted as a weighted average of $k$ ZVCV estimators where the standard ZVCV approach is a special case where $k = 1$. This can be easily extended to the vector-valued $f(\theta)$ scenario, where $\bm\beta_i$ is replaced by a $J\times T$ matrix of polynomial coefficients, where $T$ is the number of dimensions in $f(\theta)$ for multi-dimensional $f$.

Given a ZVCV design matrix $\textbf{Z}$, $\bm\beta_{i} \in \mathbb{R}^{J}, i = 1,...,k$, and a weight vector $\{w_{1},...,w_{k}\}$, the ensemble control variate estimator $\hat{I}_{\mathrm{ens}}$ for $\mathbb{E}_{\pi}[f(\theta)]$ is defined as follows:
\begin{equation}
    \hat{I}_{\mathrm{ens}} = \frac{\mathbbm{1}_{S \times 1}^\top\Big(\textbf{f}_S - \mathbf{Z}\sum_{i=1}^k w_{i}\bm\beta_{i}\Big)}{S}.
\end{equation}

\subsubsection{Estimation}
\label{section:estimation}

When $S \gg J$, $\bm\beta_i$ can be estimated using the OLS criterion, yielding the same $\hat{\bm\beta}_i$ for all $i \in \{1, \dots, k\}$. Here, the framework proposed in Section \ref{section:ecv} may be not useful as we would only need a single model (i.e., $k=1$). When $S \leq J$, the OLS solution is not identifiable. This can be avoided by constraining some polynomial coefficients of $u_{i}(\theta)$ to zero, such that the number of non-zero coefficients ($J'$) is smaller than $S$. Then, $\bm\beta_{i}$ can be estimated by solving the following constrained least squares (CLS) problem:
\begin{align}
    \big\{\hat{\alpha}_{\text{CLS},i}, \hat{\bm\beta}_{\text{CLS},i}\big\} &= \argmin_{\{\alpha_{i}, \bm\beta_{i}\}\in \mathbb{R}^{J+1}} \big\| \textbf{f}_S - \textbf{Z}\bm\beta_{i} - \alpha_{i}\big\|_2^2 \\
    \text{s.t. } \bm\beta_{i}[n] &= 0 \text{ where } \mathbb{S}_i[n] = 0 \; \; (n = 1,...,J),
\end{align}
where $\mathbb{S}_i$ denotes the selection mapping for the $i$th learner.

This is equivalent to subsetting the ZVCV design matrix $\textbf{Z}$ by column indicated by the mapping $\mathbb{S}_i$ and estimating the coefficients corresponding to the selected covariates via the OLS criterion. This variable selection task can be facilitated via LASSO \citep{leluc2021control} but can be prohibitively slow. A simpler approach, without relying on penalised regression techniques, is to randomly select $J^* < S$ columns without replacement from $\textbf{Z}$, represented by a random selection operator $\mathbb{S}_i$. Various sampling schemes can be used, which will be discussed in the next sections. If $J^* \geq J$, we simply arrive at the ZVCV estimator. We refer to our proposed method as the ensemble ZVCV algorithm (Algorithm \ref{alg:ezv}).

This random restriction acts as a form of regularisation. Each learner constrains the coefficients of the $J-J^*$ unselected covariates to zero, which restores identifiability without shrinking the coefficients of the selected covariates. With equal weights, the ensemble coefficient of the $n$th covariate, $\frac{1}{k}\sum_{i=1}^k\hat{\bm\beta}_i[n]$, equals the average of its estimates over the learners that select it, multiplied by the proportion of learners that select it. Covariates selected by only a fraction of the learners are therefore shrunk towards zero, whereas covariates included in every learner are not shrunk at all. The selection operator in Section \ref{section:selection} exploits this to retain the zero-variance property of ZVCV, which cannot be obtained directly from LASSO or ridge regression with $\lambda > 0$ because they shrink the coefficients \citep{south2023regularized}.

\begin{algorithm}[tb]
   \caption{The Ensemble ZVCV Algorithm (fixed $Q$)}
   \label{alg:ezv}
   \SetKwInput{KwData}{Input}
   \KwData{samples $\{\theta_1,...,\theta_S\}$, integrand evaluation
   $\textbf{f}_S$, gradient evaluation of log-target density $\textbf{g}_S$, the number of ensemble components $k$, the number of covariates $J$, the number of selected covariates $J^* < S$, a selection operator,
   a criteria to compute the weight vector $\textbf{w}$.}

   \lIf{$J^* \geq J$}{return the ZVCV estimate (Alg. \ref{alg:zvcv})}
   Compute the ZVCV design matrix $\mathbf{Z}$ from $\{\theta_1,...,\theta_S\}$, and $\textbf{g}_S$.

   \For{$i = 1,...,k$}{
   Initialise $\mathbb{S}_i$ via the selection operator s.t.  $\|\mathbb{S}_i\|_1 = J^*$\;
   Subsample $\textbf{Z}$ by column via $\mathbb{S}_i \Rightarrow\textbf{Z}^* \in \mathbb{R}^{S \times J^*}$\;
   Solve for $\{\hat{\alpha}_{\text{i}}, \hat{\bm\beta}_{\text{i}}\} = \argmin_{\alpha \in \mathbb{R}, \bm\beta \in \mathbb{R}^{J^*}} \big\|\textbf{Z}^*\bm\beta + \alpha - \textbf{f}_S\big\|_2^2$ \;
   }
   Compute a weight vector $\textbf{w} = \{w_1,..,w_k\}$ \;
   Return $\sum_{i=1}^kw_i\hat{\alpha}_i$.

\end{algorithm}

Note that it is not necessary to compute the complete ZVCV design matrix. If $J > kJ^*$, it is more efficient to compute each $\textbf{Z}_i$ on the fly. As in ZVCV, multiple expectations can be estimated simultaneously; our implementation solves all integrands from a single QR decomposition of each $\textbf{Z}^*$, mildly ridge-augmented ($\varepsilon = 10^{-8}$) for numerical stability. Some other details about tuning parameters and implementation will be discussed in the next sections.

\begin{remark}
\label{rem:rate}
With $d$ and $Q$ fixed, $J$ is finite while $J^*$ grows with $S$ (Section \ref{section:nselected}), so for sufficiently large $S$ every learner selects all $J$ covariates and the ensemble reduces to the ZVCV estimator with a $Q$th order polynomial. The ensemble ZVCV estimator therefore has the same asymptotic behaviour as ZVCV, including the standard $\mathcal{O}(S^{-1/2})$ convergence rate; its benefits are realised in the regime $S \leq J$, where the OLS problem for ZVCV is not identifiable.
\end{remark}

\subsection{Hyper-parameter Tuning for Ensemble ZVCV Methods}
\label{section:tuning}

Our proposed algorithm involves several hyper-parameters that need to be specified. Optimising for the \textit{best} set of hyper-parameters can be computationally expensive. Instead, we provide simple heuristics for setting these hyper-parameters that perform well in terms of statistical efficiency across various experiments in preliminary empirical assessments.

In practice, one must balance statistical and computational efficiency: if sampling is inexpensive, it is preferable to obtain additional MCMC samples. Our proposed ensemble methods are most beneficial when sampling is costly.

\subsubsection{Control Variable Selection}
\label{section:selection}

We need to determine which control variates should be selected. A naïve approach is to use simple random sampling without replacement (SRSWOR). When $Q$ and $d$ are relatively large, this random selection approach is more likely to select higher order polynomial terms (see Table \ref{tab:nrofmono}) since the number of higher-order monomials grows rapidly with polynomial order. This is inefficient when $\pi(\theta)$ is approximately Gaussian and $f(\theta)$ is a low order polynomial function (e.g. $f(\theta) = \theta$ as in posterior mean estimation). In that setting, ZVCV gives a zero-variance estimator.

\begin{table}[H]
    \renewcommand{\arraystretch}{1.0}
    \centering
    \caption{The number of monomials of order $q \in \{1,2,3,4,5\}$ for $d \in \{5,10,15\}$ and the expected polynomial order selected under the naive SRSWOR.}
    \label{tab:nrofmono}
    \begin{tabular}{rrrr}
    \toprule
    {$q$} & {$d = 5$} & {$d = 10$} & {$d = 15$}  \\ \midrule
    1 & 5 & 10  & 15  \\
    2 & 15 & 55  & 120  \\
    3 & 35 & 220  & 680 \\
    4 & 70 & 715  & 3060  \\
    5 & 126 & 2002  & 11628   \\
\midrule
   \textbf{expected} & 4.183 & 4.547 & 4.688\\
    \bottomrule
    \end{tabular}
\end{table}

To address these limitations, we propose a selection operator that guarantees polynomial exactness when $\pi(\theta)$ is Gaussian and $f(\theta)$ belongs to some classes of lower-order polynomial functions (Proposition \ref{prop:trivial}). Instead of randomly selecting $J^* < S$ columns via SRSWOR, we can choose a base polynomial order $Q^{\text{base}} < Q$ (typically, $Q^{\text{base}} = 1$ or $ Q^{\text{base}} = 2$), for which the number of base monomials is $J^{\text{base}} < J^*$, and always include the $J^{\text{base}}$ columns in $\mathbf{Z}$ corresponding to these monomials in each $\mathbf{Z}_i$ (Algorithm \ref{alg:ezv}). We then select the remaining $J^* - J^{\text{base}}$ columns from $\mathbf{Z} \backslash \mathbf{Z}^{\text{base}}$ to complete the $\mathbf{Z}_i$ matrix. Since SRSWOR over $\mathbf{Z} \backslash \mathbf{Z}^{\text{base}}$ would again favour the highest polynomial order (Table \ref{tab:nrofmono}), we use a stratified scheme: the $J^* - J^{\text{base}}$ selections are allocated approximately equally across the polynomial orders $Q^{\text{base}}+1,\ldots,Q$, and monomials are selected without replacement within each order. If all monomials of a given order are exhausted, its remaining allocation is redistributed equally among the other orders. We refer to this approach as semi-exact ensemble ZVCV. Appendix \ref{appendix:additional} compares the stratified scheme with SRSWOR over $\mathbf{Z} \backslash \mathbf{Z}^{\text{base}}$.

\begin{proposition} \label{prop:trivial} If $\pi(\theta)$ is Gaussian and $f(\theta) \in \mathcal{P}^{Q'}$, each of the $k$ components of the semi-exact ensemble ZVCV estimator with $Q^{\text{base}} \geq Q'$ will be a zero-variance estimator.
\end{proposition}

The semi-exact scheme is intended for moderate $d$, since every learner must retain all $J^{\text{base}} = \binom{d+Q^{\text{base}}}{d} - 1$ base monomials while the budget $J^* < S$ is unchanged. Alternatives for higher dimensions are discussed in Section \ref{section:other}.

\subsubsection{Number of Selected Control Variables}
\label{section:nselected}

We want $J^*$ to grow as we increase $S$. Given i.i.d. sampling, \cite{portier2019monte} show that a regression-based control variate estimator with a growing number of control variates $J^* \in \mathcal{O}(S^{1/2})$ is consistent with a non-standard super-root-$S$ convergence rate. We also expect this to hold in popular dependent sampling settings. This condition can easily be satisfied by setting $J^* = \min(c_1S, c_2\sqrt{S})$, $c_1, c_2 > 0$. By default, our algorithms set $c_1 = 0.8$ and $c_2 = 25$.

Alternatively, $J^*$ can grow linearly in $S$ as in \cite{lejeune2020implicit}. Under standard linear regression assumptions with Gaussian predictors and asymptotic subsampling conditions, a properly tuned large ensemble of bagged OLS predictors with feature randomness may exhibit an implicit regularisation equivalent to that of optimally tuned ridge regression \citep{lejeune2020implicit}. Although this has good statistical properties, these assumptions are typically violated in regression-based control variates (e.g., due to non-Gaussian control variates and misspecified linear models). Moreover, this approach can be computationally expensive when $S$ is moderately large.

\subsubsection{Combining Ensembles}
\label{section:combining}

To combine estimates from the $k$ contributing models, a simple approach is to use simple averaging (SAvg) where $w_{i} = \nicefrac{1}{k}\; (i = 1,...,k)$. As a result, the final estimate is obtained by averaging $\{\hat{\alpha}_1,...,\hat{\alpha}_k\}$. Another approach is based on minimising the variance of the control variate estimator (Proposition \ref{prop:error-bounds}). If weights are bounded, the optimal weights $\mathbf{w}^{\mathrm{opt}}$ can be obtained by solving the following quadratic programming problem:

\begin{align}
\label{eqn:quadprog}
\mathbf{w}^{\mathrm{opt}}
&= \arg\min_{\mathbf{w} \in [0,1]^k}
\mathbf{w}^\top \, \mathlarger{\Sigma}_{\hat{\alpha}} \, \mathbf{w} \\
&\text{s.t.} \quad \sum_{i=1}^k w_i = 1,
\end{align}

where $\mathlarger{\Sigma}_{\hat{\alpha}}$ denotes the covariance matrix of $\{\hat{\alpha}_1,...,\hat{\alpha}_k\}$. Obtaining an unbiased estimator of  $\mathlarger{\Sigma}_{\hat{\alpha}}$ can be mathematically challenging, especially for dependent sampling settings. However, we can replace the unknown covariance matrix $\mathlarger{\Sigma}_{\hat{\alpha}}$ with a proxy estimator. A simple and computationally efficient option is to use the sample covariance matrix of the leave-one-out (LOO) residuals of the $k$ regression tasks. We use LOO rather than in-sample residuals because the in-sample residuals of each learner are shrunk towards zero by overfitting, increasingly so as $J^*$ increases. Exploiting the linear structure of OLS, the LOO residuals can be computed from the in-sample residuals and leverages of each fit, without refitting any regression (Appendix \ref{appendix:combining}). We refer to this approach as QP$_+$, as it is a quadratic programme with non-negative weights, analogous to Markowitz portfolio optimisation \citep{markowitz2008portfolio}.

If weights are unbounded, i.e.\ only constrained to sum to one, the optimal weights can be obtained in closed form. We refer to this variant as $\mathrm{QP}_{\pm}$. In our experiments it generally performed worse than QP$_+$, since negative weights can amplify the estimation error in the proxy covariance matrix, so we report its results only in Appendix \ref{appendix:additional}.

\subsubsection{Ensemble Size}
\label{section:size}

For SAvg-based approaches, we expect greater variance reduction as $k$ increases. This is analogous to the behaviour of random forests, where the algorithm performance does not degrade (i.e., exhibit overfitting) with the number of trees but instead improves (i.e., decreasing variance, \citealt[p.588]{hastie2009elements}). We find $k = 25$ performs well across many examples considered in terms of statistical efficiency.

Unlike the SAvg-based approaches, QP$_+$-based approaches can exhibit overfitting as $k$ increases because weights are estimated. Here we also suggest $k = 25$.

\subsubsection{Other Potential Choices}
\label{section:other}

Our proposed methods do not scale well computationally as $d$ increases. Various more computationally efficient selection operators exist, albeit at the expense of statistical efficiency. One approach is to fix the monomials corresponding to non‐interaction terms and perform random sampling on the others. Another possibility, which we investigate in Experiment 3, is to select parameters rather than monomials to construct a smaller ZVCV design matrix, analogous to the \textit{a priori} ZVCV method in \cite{south2023regularized}. In \textit{a priori} ZVCV, $u(\theta)$ is a $Q$th order polynomial in a subset of the dimensions of $\theta$ that are expected to be relevant to $f(\theta)$. For an ensemble version, when $f(\theta)$ depends only on some anchor dimensions, these dimensions can be included in every learner, and each learner uses all monomials of order up to $Q$ in the anchor dimensions together with a learner-specific subset of the remaining dimensions. We refer to this approach as ensemble \textit{a priori} ZVCV. We do not use a fixed rule for choosing $J^*$ under the \textit{a priori} selection strategy. As this strategy is tailored to a specific integrand, $J^*$ is determined by the dimensions deemed relevant to that integrand and the polynomial order, subject to $J^* < S$. Because the selection is performed separately for each expectation, smaller learners may be preferable when many expectations are estimated. Under the \textit{a priori} selection strategy, the constituent learners may have substantially different variances, making variance-based weighting (Section \ref{section:combining}) particularly beneficial.

Dimensional reduction techniques, such as principal component analysis and random projection \citep{johnson1984extensions}, do not select control variates directly but can instead be used to construct lower-dimensional control variates. We do not investigate these alternatives here.

\section{Simulation Study}
\label{section:sstudy}
\subsection{Simulation Methodology}
\label{section:simMed}
We have conducted a simulation study to evaluate the proposed methods against established control variate techniques in the literature. The specific methods include:

\begin{itemize}
    \item[--] Semi-exact ensemble ZVCVs methods (Experiments 1 and 2): $\mathrm{eZV}_{k,\mathrm{SAvg}}$ and $\mathrm{eZV}_{k,\mathrm{QP}_{+}}$ where $k \in \{1,25,50\}$, $Q_{\text{base}} = \argmax_{q = 1:2}{{d+q\choose d}<S}$, and  $Q_{\max} = 5$. For $k=1$, both methods reduce to a single learner, which we denote $\mathrm{eZV}_{1}$. We follow the heuristics presented in the previous sections for setting hyper-parameters.
    \item[--] Ensemble \textit{a priori} ZVCV methods (Experiment 3): $\mathrm{eZV}_{m_{\max},\mathrm{SAvg}}$ and $\mathrm{eZV}_{m_{\max},\mathrm{QP}_{+}}$ (see Section \ref{exp:3}).
    \item[--] ZVCV with OLS and $Q$-th order polynomial: $\text{ZV}_{Q}$, where ${Q+d\choose Q} < S$, $Q \in \{1,2,3,4,5\}$.
    \item[--] \textit{A priori} ZVCV with a $Q$-th order polynomial in $m$ selected dimensions (Experiment 3): $\text{a-ZV}_{m,Q}$ (see Section \ref{exp:3}).
    \item[--] Regularised ZVCV with LASSO or ridge regression and $Q$-th order polynomial: $\ell$-$\text{ZV}_Q$ and $r$-$\text{ZV}_Q$, respectively; $\lambda$ is selected using 10-fold cross validation, $Q \in \{1,2,3,4,5\}$ ($Q \in \{1,2\}$ in Experiment 3).
    \item[--] Control functionals with Gaussian-kernel via median-tuned heuristic: $\text{CF}$ (see \cite{garreau2017large} for details regarding the median-tuned heuristic).
    \item[--] Semi-exact control functionals with a Gaussian kernel via median-tuned heuristic: $\text{SECF}_Q$ where ${Q+d\choose Q} < S$, $Q \in \{1,2\}$.
\end{itemize}

The ensemble ZVCV methods used in the simulation study are available in the GitHub repository \href{https://github.com/edelweiss611428/ZVengine}{\texttt{edelweiss611428/ZVengine}}. The competing methods are implemented in the \texttt{ZVCV} package \citep{southzvcvpackage2022}. The regularised ZVCV methods utilise penalised regression solvers in the \texttt{glmnet} package \citep{friedman2021package}. Here, the sequence of $\lambda$ is optimised for statistical efficiency, which is the default option in the R package \texttt{ZVCV}, instead of relying on the automatically generated sequences from \texttt{glmnet}.

For each experiment, we evaluate a method $\mathcal{M}$ via statistical efficiency (SE) and overall efficiency (OE) with respect to the baseline approach -- vanilla Monte Carlo. For each integrand, these quantities can be estimated as follows:

\begin{gather}
    \label{eqn:vrf}
    \reallywidehat{\text{SE}(\mathcal{M})} = \frac{\reallywidehat{\text{MSE}[\hat{I}_{\text{MC}}]}}{\reallywidehat{\text{MSE}[\mathcal{M}]}}, \\
    \reallywidehat{\text{OE}(\mathcal{M})} = \reallywidehat{\text{SE}(\mathcal{M})} \times \frac{\reallywidehat{\text{runtime}(\hat{I}_{\text{MC}})}}{\reallywidehat{\text{runtime}(\mathcal{M})} + \reallywidehat{\text{runtime}(\hat{I}_{\text{MC}})}},
\end{gather}

where MSE is the mean-squared error of a method, estimated using a gold standard. For Experiments 1 and 2, we use a gold standard $\text{ZV}_3$ estimate, where a sample of size $10^5$ (with $10^5$ warm-up iterations) is used to estimate $\text{ZV}_3$'s regression coefficients, and another sample of similar size, obtained independently of the previous one with the same number of warm-up iterations, is used to compute the $\text{ZV}_3$ estimate (similar to the ZVCV approach analysed in \cite{mira2013zero}). Here, the $\text{ZV}_3$ estimator could be approximately unbiased and should have a lower MSE than a vanilla estimate based on a single sample of size $2\times10^5$, for example. For Experiment 3, the gold standard is obtained from 10 independent chains, each with $10^5$ samples after warm-up.

If there are $T$ expectations to be estimated, $\reallywidehat{\text{SE}(\mathcal{M})}$ and $\reallywidehat{\text{OE}(\mathcal{M})}$ will be averaged over $T$ tasks.

\subsection{Experiments}
\label{section:exp_details}

Table~\ref{tab:examples} provides brief descriptions of the experiments in the simulation study. Detailed descriptions are shown in the next sections. We use \texttt{rstan} \citep{carpenter2017stan} to sample from the posterior distribution. Sampler configurations are shown in Table~\ref{tab:examples}. In Experiments 1 and 2, we vary the sample size $S \in \{100, 300, 1000, 3000\}$; in Experiment 3, $S \in \{300, 1000\}$. The number of \texttt{warmup} iterations is 1000. We only use a single MCMC chain. For each sample size, each experiment is repeated 100 times.

\begin{table}[H]
    \renewcommand{\arraystretch}{1.0}
    \centering
    \caption{Brief descriptions of the experiments in the simulation study, where $d$ is the number of dimensions of $\Theta$.}
    \label{tab:examples}
    \resizebox{1\textwidth}{!}{
    \begin{tabular}{cccccc} \toprule
    {\textbf{id}} & {\textbf{Experiment}}  & {\textbf{d}} & {\textbf{Task}} & {\textbf{Sampler configurations}} & {\textbf{ODE solver}} \\ \midrule
    1. & Lotka-Volterra & 8 & $\mathbb{E}[\theta]$  & \texttt{adapt\_delta = 0.99},  \texttt{max\_treedepth = 15} & \texttt{rk45}  \\
    2. & Friberg-Karlsson & 11 & $\mathbb{E}[\theta]$  & \texttt{adapt\_delta = 0.95}, custom initialisation & \texttt{rk45} \\
    3. & Log-Gaussian Cox process & 64 & $\mathbb{E}[\theta]$  & default & -- \\
    \bottomrule
    \end{tabular}
    }
\end{table}

\subsubsection{$\textbf{Experiment 1: }$ Lotka-Volterra Model}
\label{exp:1}
Experiment 1 considers a Lotka–Volterra model \citep{lotka1910contribution, volterra1926variazioni} for modelling population dynamics of two species: snowshoe hares (prey) and Canadian lynxes (predator), given \textit{observed} predator counts $y_{v}[t]$ and prey counts $y_{u}[t]$ ($t \in \{0, 1, \ldots, 20\}$) in the $\texttt{hudson\_lynx\_hare}$ dataset from \texttt{posteriordb} \citep{magnusson2024posteriordb}. The model involves a system of first-order ordinary differential equations (ODEs) describing the population dynamics of the predator–prey pair. Particularly,
\begin{align}
\frac{du}{dt} &= (\alpha - \beta v[t])\, u[t] \label{eqn:pred} \\
\frac{dv}{dt} &= (-\gamma + \delta u[t])\, v[t] \label{eqn:prey},
\end{align}
where $u[t]$ and $v[t]$ denote the \textit{actual} population sizes of the prey and predator species at time $t \in \{0, 1, \ldots, 20\}$, respectively, and $\alpha, \beta, \gamma$, and $\delta$ are positive ecological parameters. Given fixed parameters $\{u(0), v(0), \alpha, \beta, \gamma, \delta\}$, denoted as $\theta$, the \textit{observed} predator and prey counts are modelled using the following noise processes:
\begin{align}
\label{eqn:noise_u}
y_{u}[t] | \theta &\sim \text{LogNormal}\big(\log(u[t]) | \theta, \sigma_u^2\big) \\
y_{v}[t] | \theta &\sim \text{LogNormal}\big(\log (v[t])| \theta, \sigma_v^2\big) \label{eqn:noise_v}
\end{align}

The following prior distributions are used:
\begin{align*}
\alpha, \gamma  &\sim \mathcal{N}(1, 0.5^2) \\
\beta, \delta  &\sim \mathcal{N}(0.05, 0.05^2) \\
u(0), v(0) &\sim \mathrm{LogNormal}(\log 10, 1^2)\\
\sigma_u, \sigma_v  &\sim \mathrm{LogNormal}(-1, 1^2)
\end{align*}

The \texttt{Stan} code for Experiment 1 is available in \texttt{lotka\_volterra.stan}  from \texttt{posteriordb}. To solve the system of ODEs, we use the \texttt{integrate\_ode\_rk45} solver with default tolerance settings. One replication is discarded where the Stan sampler gets stuck and fails to explore the posterior distribution. Here, it takes an unreasonable amount of time to finish. The discarded sample's log posterior densities are also generally smaller than these of the other samples.

We aim to estimate posterior means of $u(0), v(0), \alpha, \beta, \gamma, \delta, \sigma_u$, and $\sigma_v$.

\subsubsection{\textbf{Experiment 2: } Friberg-Karlsson model}
\label{exp:2}

\begin{figure}[htbp]
    \centering
    \includegraphics[width=0.7\textwidth]{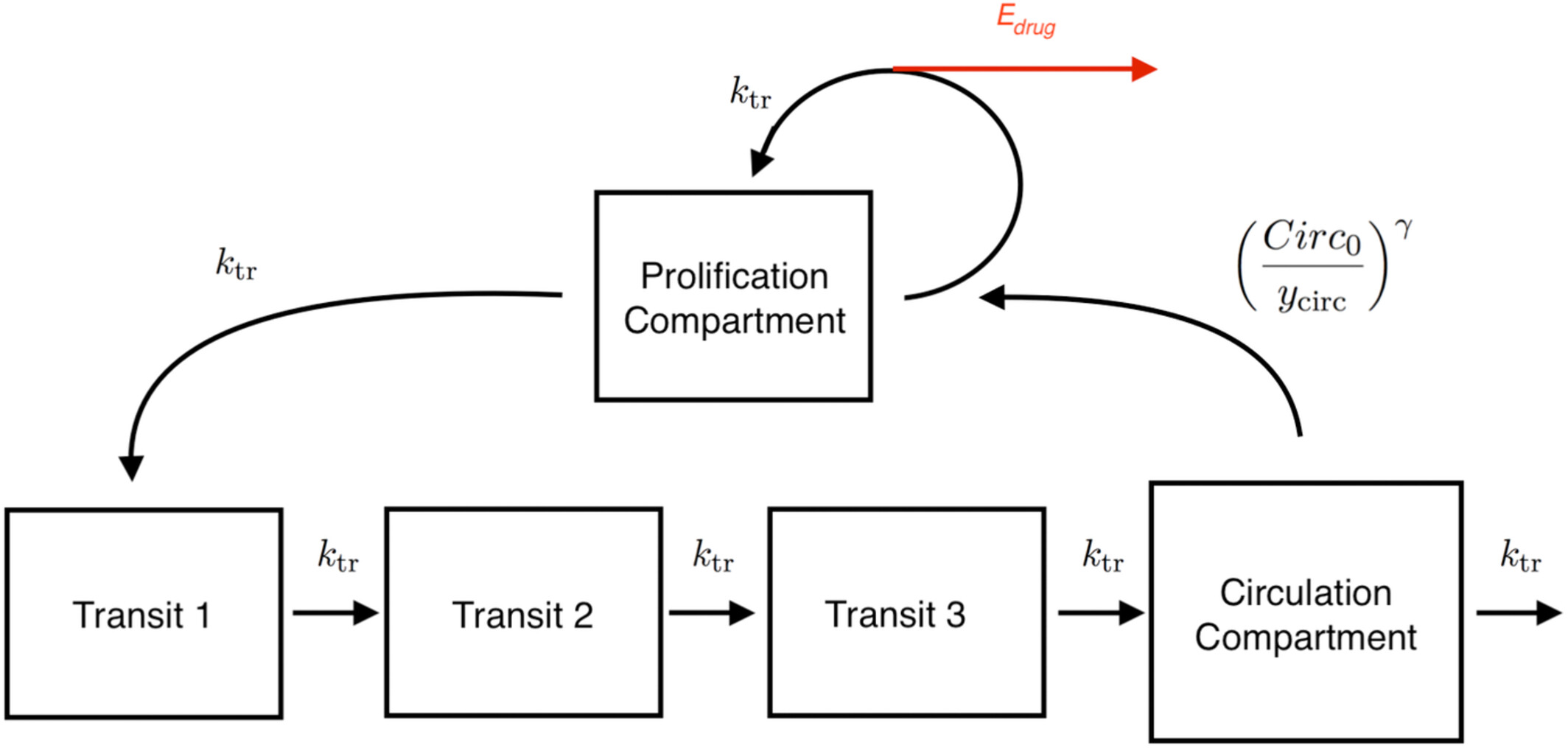}
    \caption{Semi-mechanistic model of drug-induced myelosuppression \citep{margossian2022flexible}. \textit{Prol}: Stem cell compartment; \textit{Transit 1}: Maturation compartment 1; \textit{Transit 2}: Maturation compartment 2; \textit{Transit 3}: Maturation compartment 3; \textit{Circ}: Circulating neutrophil compartment \citep{latz2006semimechanistic}}
    \label{fig:myfigure}
\end{figure}

Experiment 2 considers a Friberg–Karlsson model \citep{friberg2003mechanistic, latz2006semimechanistic, quartino2011semi} for studying drug-induced myelosuppression, a condition involving reduced bone marrow activity resulting in decreased neutrophil cell production. This is a semi-mechanistic pharmacokinetic – pharmacodynamic (PK–PD) model describing a delayed feedback mechanism between proliferative neutrophil cells and mature neutrophil cells in circulation that maintains the absolute neutrophil count (ANC) near a baseline level, taking into account the drug effect disturbing the regulatory system \citep{margossian2022flexible}. The PD diagram is shown in Figure \ref{fig:myfigure} \citep{margossian2022flexible}, and the corresponding model likelihood is:
\begin{align}
y_{ANC}[t]|\theta \sim  \text{LogNormal}\big(Circ[t]|\theta, \sigma^2_{ANC}\big),
\end{align}
where $\theta$ denotes model parameters, and $Circ[t]$ is obtained by solving the following system of ODEs \citep{latz2006semimechanistic, margossian2022flexible}:
\begin{align*}
\frac{d\,\textit{Prol}}{dt} &= k_{\textit{Prol}} \cdot \textit{Prol} \cdot (1 - E_{\textit{Drug}}) \cdot \left( \frac{\textit{Circ}_0}{\textit{Circ}} \right)^{\gamma} - k_{\textit{tr}} \cdot \textit{Prol} \\
\frac{d\,\textit{Transit}_1}{dt} &= k_{\textit{tr}} \cdot \textit{Prol} - k_{\textit{tr}} \cdot \textit{Transit}_1 \\
\frac{d\,\textit{Transit}_2}{dt} &= k_{\textit{tr}} \cdot \textit{Transit}_1 - k_{\textit{tr}} \cdot \textit{Transit}_2 \\
\frac{d\,\textit{Transit}_3}{dt} &= k_{\textit{tr}} \cdot \textit{Transit}_2 - k_{\textit{tr}} \cdot \textit{Transit}_3 \\
\frac{d\,\textit{Circ}}{dt} &= k_{\textit{tr}} \cdot \textit{Transit}_3 - k_{\textit{Circ}} \cdot \textit{Circ}
\end{align*}

Here, $E_{Drug} = \min(\alpha\hat{c}, 1)$, where $\hat{c}$ is the drug concentration calculated from the PK model \citep{margossian2022flexible}, which is not shown in the diagram (see \citealt{latz2006semimechanistic} for detailed descriptions of the model). The prior distributions for the 11 model parameters are taken from \citet{margossian2022flexible}.

The \texttt{Stan} code for Experiment 2 is based on the \texttt{neutropenia} example in the GitHub repository \texttt{n000001/Stan-PKPD-examples} \citep{baldy2023hierarchical}. Our task is to estimate the posterior means of the 11 model parameters.

\subsubsection{\textbf{Experiment 3: } Log-Gaussian Cox Process}
\label{exp:3}

Experiment 3 considers a log-Gaussian Cox process \citep{moller1998log, heng2019unbiased} for the \texttt{finpines} dataset from \texttt{spatstat}, which has previously been used to assess Stein-based control variates \citep{south2023regularized}. The observed pine-sapling locations are normalised to the unit square and partitioned into an $n\times n$ regular grid with $n = 8$, giving $d = 64$. Let $y_{ij}$ denote the number of observed points in grid cell $(i,j)$. Conditional on the latent log-intensity $\theta_{ij}$, the cell counts are independent and satisfy
\begin{align}
y_{ij}|\theta_{ij}
&\sim
\mathrm{Poisson}\left(\frac{\exp(\theta_{ij})}{n^2}\right),
\qquad i,j=1,\ldots,n.
\end{align}
The latent log-intensity has a Gaussian process prior $\bm{\theta} \sim \mathcal{N}(\mu\mathbbm{1},\bm{\Sigma}_{\bm\theta})$, where
\begin{align}
\mu
&=
\log(126)-\frac{s^2}{2},\\
\bm{\Sigma}_{\bm\theta}[(i,j),(i',j')]
&=
s^2
\exp\left\{
-\frac{
\sqrt{(i-i')^2+(j-j')^2}
}{
n\beta
}
\right\},
\end{align}
with $s^2=1.91$ and $\beta=1/33$. The model is implemented in \texttt{Stan}, and we estimate the posterior means of the $d=64$ latent log-intensity parameters.

As $d = 64$ is too large for the semi-exact scheme (e.g., $J^{\text{base}} = 2144$ for $Q^{\text{base}} = 2$), we use ensemble \textit{a priori} ZVCV (Section \ref{section:other}), with the target cell as the anchor dimension. For each target cell $c$, let $\mathcal{N}_m(c)$ denote the $m$ grid cells nearest to $c$, ranked by the Euclidean distance between grid coordinates (ties broken by cell index), so that $c \in \mathcal{N}_1(c) \subset \mathcal{N}_2(c) \subset \cdots$. The $m$th learner is an \textit{a priori} ZVCV with a fourth-order polynomial in the dimensions in $\mathcal{N}_m(c)$, which has $\binom{m+4}{4} - 1$ covariates. The ensemble combines the learners $m = 1, \ldots, m_{\max}(S)$, where $m_{\max}(S) = \min\big(9, \max\{m : \binom{m+4}{4} < S\}\big)$ ensures that each learner has fewer covariates than samples, giving $m_{\max} = 6$ for $S = 300$ and $m_{\max} = 9$ for $S = 1000$. As baselines, we also report the two extreme learners that the ensemble combines: $\text{a-ZV}_{1,4}$, which uses the target cell only, and $\text{a-ZV}_{m_{\max},4}$, which uses the largest feasible neighbourhood.

\subsection{Results}

In this section, we comment on the results of the experiments described in Section \ref{section:exp_details}. All performance metrics are averaged over 100 independent replications (99 for Experiment 1, where one replication was discarded).

\subsubsection{Experiment 1: Lotka-Volterra Model}
\begin{table}[t]
    \renewcommand{\arraystretch}{1}
    \centering
    \caption{Results of Experiment 1: Lotka-Volterra model}
    \label{table:lv}
    \resizebox{1\textwidth}{!}{
\begin{tabular}{ccccccccc}
\toprule
& \multicolumn{4}{c}{Statistical efficiency} & \multicolumn{4}{c}{Overall efficiency} \\
\cmidrule(r){2-5}\cmidrule(l){6-9}
Method & $S = 100$ & $S = 300$ & $S = 1000$ & $S = 3000$ & $S = 100$ & $S = 300$ & $S = 1000$ & $S = 3000$ \\
\midrule
$\text{ZV}_1$ & $9.29$ & $8.99$ & $8.54$ & $7.62$ & $9.29$ & $8.99$ & $8.54$ & $7.62$ \\
$\text{ZV}_2$ & $6.43\times10^{1}$ & $1.06\times10^{2}$ & $1.28\times10^{2}$ & $1.22\times10^{2}$ & $6.43\times10^{1}$ & $1.06\times10^{2}$ & $1.28\times10^{2}$ & $1.22\times10^{2}$ \\
$\text{ZV}_3$ & -- & $1.11\times10^{3}$ & $2.11\times10^{3}$ & $2.10\times10^{3}$ & -- & $1.11\times10^{3}$ & $2.10\times10^{3}$ & $2.10\times10^{3}$ \\
$\text{ZV}_4$ & -- & -- & $1.44\times10^{4}$ & $1.59\times10^{4}$ & -- & -- & $1.44\times10^{4}$ & $1.58\times10^{4}$ \\
$\text{ZV}_5$ & -- & -- & -- & $1.40^{\dagger}$ & -- & -- & -- & $1.31^{\dagger}$ \\
\midrule
$r\text{-}\text{ZV}_1$ & $9.02$ & $8.89$ & $8.49$ & $7.59$ & $8.96$ & $8.83$ & $8.43$ & $7.54$ \\
$r\text{-}\text{ZV}_2$ & $2.81\times10^{1}$ & $6.62\times10^{1}$ & $9.74\times10^{1}$ & $9.63\times10^{1}$ & $2.67\times10^{1}$ & $6.14\times10^{1}$ & $9.13\times10^{1}$ & $9.24\times10^{1}$ \\
$r\text{-}\text{ZV}_3$ & $3.14\times10^{1}$ & $9.05\times10^{1}$ & $1.21\times10^{2}$ & $1.48\times10^{2}$ & $2.47\times10^{1}$ & $6.67\times10^{1}$ & $9.48\times10^{1}$ & $1.27\times10^{2}$ \\
$r\text{-}\text{ZV}_4$ & $2.78\times10^{1}$ & $9.67\times10^{1}$ & $1.40\times10^{2}$ & $1.70\times10^{2}$ & $1.25\times10^{1}$ & $3.64\times10^{1}$ & $6.16\times10^{1}$ & $9.51\times10^{1}$ \\
$r\text{-}\text{ZV}_5$ & $2.59\times10^{1}$ & $8.75\times10^{1}$ & $1.38\times10^{2}$ & $1.77\times10^{2}$ & $1.98\times10^{1}$ & $4.32\times10^{1}$ & $3.91\times10^{1}$ & $1.53\times10^{1}$ \\
\midrule
$\ell\text{-}\text{ZV}_1$ & $8.94$ & $8.83$ & $8.48$ & $7.59$ & $8.88$ & $8.77$ & $8.43$ & $7.54$ \\
$\ell\text{-}\text{ZV}_2$ & $3.30\times10^{1}$ & $6.87\times10^{1}$ & $9.72\times10^{1}$ & $9.04\times10^{1}$ & $3.05\times10^{1}$ & $6.25\times10^{1}$ & $9.02\times10^{1}$ & $8.62\times10^{1}$ \\
$\ell\text{-}\text{ZV}_3$ & $3.35\times10^{1}$ & $8.68\times10^{1}$ & $1.18\times10^{2}$ & $1.33\times10^{2}$ & $2.80\times10^{1}$ & $6.88\times10^{1}$ & $9.91\times10^{1}$ & $1.19\times10^{2}$ \\
$\ell\text{-}\text{ZV}_4$ & $3.10\times10^{1}$ & $9.05\times10^{1}$ & $1.34\times10^{2}$ & $1.52\times10^{2}$ & $2.03\times10^{1}$ & $5.51\times10^{1}$ & $8.28\times10^{1}$ & $1.05\times10^{2}$ \\
$\ell\text{-}\text{ZV}_5$ & $2.75\times10^{1}$ & $8.81\times10^{1}$ & $1.33\times10^{2}$ & $1.59\times10^{2}$ & $2.26\times10^{1}$ & $5.27\times10^{1}$ & $4.92\times10^{1}$ & $2.30\times10^{1}$ \\
\midrule
$\text{CF}$ & $1.52$ & $3.37$ & $9.33$ & $1.91\times10^{1}$ & $1.52$ & $3.36$ & $9.27$ & $1.76\times10^{1}$ \\
$\text{SECF}_1$ & $1.45\times10^{1}$ & $1.81\times10^{1}$ & $1.06\times10^{1}$ & $7.84$ & $1.45\times10^{1}$ & $1.81\times10^{1}$ & $1.06\times10^{1}$ & $7.26$ \\
$\text{SECF}_2$ & $8.91\times10^{1}$ & $1.47\times10^{2}$ & $2.28\times10^{2}$ & $3.65\times10^{2}$ & $8.91\times10^{1}$ & $1.47\times10^{2}$ & $2.27\times10^{2}$ & $3.37\times10^{2}$ \\
\midrule
$\bm{\mathrm{eZV}_{1}}$ & $4.50\times10^{1}$ & $6.75\times10^{2}$ & $3.12\times10^{4}$ & \textcolor{red}{$\bm{4.33\times10^{4}}$} & $4.50\times10^{1}$ & $6.75\times10^{2}$ & $3.06\times10^{4}$ & \textcolor{red}{$\bm{3.98\times10^{4}}$} \\
$\bm{\mathrm{eZV}_{25,\mathrm{SAvg}}}$ & \textcolor{red}{$\bm{1.08\times10^{2}}$} & $2.03\times10^{3}$ & $5.03\times10^{4}$ & \textcolor{red}{$\bm{4.33\times10^{4}}$} & \textcolor{red}{$\bm{1.08\times10^{2}}$} & $2.00\times10^{3}$ & \textcolor{red}{$\bm{3.37\times10^{4}}$} & \textcolor{red}{$\bm{3.98\times10^{4}}$} \\
$\bm{\mathrm{eZV}_{50,\mathrm{SAvg}}}$ & $1.08\times10^{2}$ & \textcolor{red}{$\bm{2.17\times10^{3}}$} & \textcolor{red}{$\bm{5.13\times10^{4}}$} & \textcolor{red}{$\bm{4.33\times10^{4}}$} & $1.08\times10^{2}$ & \textcolor{red}{$\bm{2.10\times10^{3}}$} & $2.58\times10^{4}$ & \textcolor{red}{$\bm{3.98\times10^{4}}$} \\
\midrule
$\bm{\mathrm{eZV}_{25,\mathrm{QP}_{+}}}$ & $8.61\times10^{1}$ & $1.76\times10^{3}$ & $4.76\times10^{4}$ & \textcolor{red}{$\bm{4.33\times10^{4}}$} & $8.61\times10^{1}$ & $1.73\times10^{3}$ & $3.18\times10^{4}$ & \textcolor{red}{$\bm{3.98\times10^{4}}$} \\
$\bm{\mathrm{eZV}_{50,\mathrm{QP}_{+}}}$ & $9.59\times10^{1}$ & $2.09\times10^{3}$ & $4.95\times10^{4}$ & \textcolor{red}{$\bm{4.33\times10^{4}}$} & $9.58\times10^{1}$ & $2.01\times10^{3}$ & $2.49\times10^{4}$ & \textcolor{red}{$\bm{3.98\times10^{4}}$} \\
\bottomrule
\end{tabular}
}

{\footnotesize $^{\dagger}$ At $S=3000$, the fifth-order basis has $J=1286$
covariates, so the ensemble methods select it in full and coincide with $\text{ZV}_5$ up to the linear solver: $\text{ZV}_5$ uses \texttt{lm()} in the \texttt{ZVCV} package at its default tolerance, whereas our implementation uses a QR decomposition with a mild ridge augmentation ($\varepsilon=10^{-8}$). The reported $\text{ZV}_5$ efficiency therefore reflects the numerical conditioning of the solver rather than the polynomial basis.}

\end{table}

In Experiment 1, sampling from the posterior distribution is relatively expensive. For most methods, the overall efficiency is primarily dominated by the statistical efficiency. The exceptions are the regularised ZVCV methods with higher polynomial orders and, at $S \geq 1000$, the ensemble methods, for which the fitting cost is no longer negligible relative to sampling. Our proposed methods achieve the best statistical and overall efficiency at every sample size. Their statistical efficiency improves as $k$ increases from 1 to 25, with minimal differences between $k = 25$ and $k = 50$. At $S = 1000$, the best ensemble improves on the best non-ensemble method, $\text{ZV}_4$, by a factor of roughly $3.6$ in statistical efficiency, and on the best regularised ZVCV method by a factor of more than 300. Given the default options, the SAvg-based method slightly outperforms the QP$_+$-based one at every sample size at which the ensembles differ. At $S = 3000$, all ensembles coincide: with $d = 8$ and $Q = 5$, the full basis has $J = 1286$ covariates, while $J^* = \min(c_1S, c_2S^{1/2}) = 1369$, so every learner selects the entire basis.

ZVCV consistently underperforms compared to our proposed approaches; its performance improves as $Q$ increases from 1 to 4, while the poor performance of $\text{ZV}_5$ at $S = 3000$ is due to the numerical conditioning of the linear solver (see the footnote of Table \ref{table:lv}). This may explain why our proposed approaches work well in this example, as they have earlier access to higher-order monomials. The statistical efficiencies of regularised ZVCV approaches are improved as we increase the polynomial order; however, they do not benefit much from increasing the polynomial order beyond $Q = 3$ and become noticeably slower compared to the other methods. For example, at $S = 1000$, the statistical efficiency of $\ell$-$\text{ZV}_Q$ improves only from $1.18\times10^{2}$ to $1.33\times10^{2}$ as $Q$ increases from 3 to 5, while its overall efficiency declines from $9.91\times10^{1}$ to $4.92\times10^{1}$. CF and $\text{SECF}_1$ are among the least statistically efficient methods, but this has been improved upon by $\text{SECF}_2$.

\subsubsection{Experiment 2: Friberg-Karlsson Model}

\begin{table}[t]
    \renewcommand{\arraystretch}{1}
    \centering
    \caption{Results of Experiment 2: Friberg-Karlsson model}
    \label{table:fk}
    \resizebox{1\textwidth}{!}{
\begin{tabular}{ccccccccc}
\toprule
& \multicolumn{4}{c}{Statistical efficiency} & \multicolumn{4}{c}{Overall efficiency} \\
\cmidrule(r){2-5}\cmidrule(l){6-9}
Method & $S = 100$ & $S = 300$ & $S = 1000$ & $S = 3000$ & $S = 100$ & $S = 300$ & $S = 1000$ & $S = 3000$ \\
\midrule
$\text{ZV}_1$ & $7.91$ & $7.53$ & $9.19$ & $9.13$ & $7.91$ & $7.53$ & $9.19$ & $9.13$ \\
$\text{ZV}_2$ & $4.01\times10^{1}$ & $1.74\times10^{2}$ & $2.17\times10^{2}$ & $2.10\times10^{2}$ & $4.01\times10^{1}$ & $1.74\times10^{2}$ & $2.17\times10^{2}$ & $2.10\times10^{2}$ \\
$\text{ZV}_3$ & -- & -- & $6.69\times10^{3}$ & $6.43\times10^{3}$ & -- & -- & $6.69\times10^{3}$ & $6.43\times10^{3}$ \\
$\text{ZV}_4$ & -- & -- & -- & \textcolor{red}{$\bm{3.98\times10^{4}}$} & -- & -- & -- & \textcolor{red}{$\bm{3.97\times10^{4}}$} \\
$\text{ZV}_5$ & -- & -- & -- & -- & -- & -- & -- & -- \\
\midrule
$r\text{-}\text{ZV}_1$ & $7.86$ & $7.49$ & $9.16$ & $9.12$ & $7.85$ & $7.48$ & $9.15$ & $9.11$ \\
$r\text{-}\text{ZV}_2$ & $1.84\times10^{1}$ & $4.44\times10^{1}$ & $7.86\times10^{1}$ & $8.59\times10^{1}$ & $1.78\times10^{1}$ & $4.26\times10^{1}$ & $7.62\times10^{1}$ & $8.45\times10^{1}$ \\
$r\text{-}\text{ZV}_3$ & $2.14\times10^{1}$ & $6.38\times10^{1}$ & $1.08\times10^{2}$ & $1.18\times10^{2}$ & $1.74\times10^{1}$ & $4.72\times10^{1}$ & $8.40\times10^{1}$ & $1.02\times10^{2}$ \\
$r\text{-}\text{ZV}_4$ & $2.05\times10^{1}$ & $7.04\times10^{1}$ & $1.26\times10^{2}$ & $1.57\times10^{2}$ & $1.88\times10^{1}$ & $5.17\times10^{1}$ & $5.46\times10^{1}$ & $2.99\times10^{1}$ \\
$r\text{-}\text{ZV}_5$ & $1.60\times10^{1}$ & $6.42\times10^{1}$ & $1.42\times10^{2}$ & $2.09\times10^{2}$ & $1.33\times10^{1}$ & $3.43\times10^{1}$ & $2.71\times10^{1}$ & $1.62\times10^{1}$ \\
\midrule
$\ell\text{-}\text{ZV}_1$ & $7.85$ & $7.45$ & $9.11$ & $9.09$ & $7.83$ & $7.44$ & $9.10$ & $9.08$ \\
$\ell\text{-}\text{ZV}_2$ & $2.48\times10^{1}$ & $5.17\times10^{1}$ & $8.49\times10^{1}$ & $8.87\times10^{1}$ & $2.39\times10^{1}$ & $4.98\times10^{1}$ & $8.30\times10^{1}$ & $8.77\times10^{1}$ \\
$\ell\text{-}\text{ZV}_3$ & $2.46\times10^{1}$ & $6.35\times10^{1}$ & $1.02\times10^{2}$ & $1.14\times10^{2}$ & $2.18\times10^{1}$ & $5.41\times10^{1}$ & $8.91\times10^{1}$ & $1.06\times10^{2}$ \\
$\ell\text{-}\text{ZV}_4$ & $2.50\times10^{1}$ & $7.32\times10^{1}$ & $1.22\times10^{2}$ & $1.31\times10^{2}$ & $2.37\times10^{1}$ & $6.13\times10^{1}$ & $6.53\times10^{1}$ & $3.73\times10^{1}$ \\
$\ell\text{-}\text{ZV}_5$ & $2.44\times10^{1}$ & $7.61\times10^{1}$ & $1.24\times10^{2}$ & $1.44\times10^{2}$ & $2.24\times10^{1}$ & $5.71\times10^{1}$ & $4.06\times10^{1}$ & $2.81\times10^{1}$ \\
\midrule
$\text{CF}$ & $1.40$ & $2.81$ & $6.73$ & $2.03\times10^{1}$ & $1.40$ & $2.81$ & $6.72$ & $2.00\times10^{1}$ \\
$\text{SECF}_1$ & $1.08\times10^{1}$ & $1.51\times10^{1}$ & $2.81\times10^{1}$ & $3.52\times10^{1}$ & $1.08\times10^{1}$ & $1.51\times10^{1}$ & $2.81\times10^{1}$ & $3.48\times10^{1}$ \\
$\text{SECF}_2$ & \textcolor{red}{$\bm{4.65\times10^{1}}$} & $2.39\times10^{2}$ & $4.37\times10^{2}$ & $6.83\times10^{2}$ & \textcolor{red}{$\bm{4.65\times10^{1}}$} & $2.39\times10^{2}$ & $4.37\times10^{2}$ & $6.75\times10^{2}$ \\
\midrule
$\bm{\mathrm{eZV}_{1}}$ & $3.24\times10^{1}$ & $1.47\times10^{2}$ & $3.47\times10^{3}$ & $1.76\times10^{4}$ & $3.24\times10^{1}$ & $1.47\times10^{2}$ & $3.46\times10^{3}$ & $1.74\times10^{4}$ \\
$\bm{\mathrm{eZV}_{25,\mathrm{SAvg}}}$ & $4.28\times10^{1}$ & $4.28\times10^{2}$ & $1.08\times10^{4}$ & $2.94\times10^{4}$ & $4.28\times10^{1}$ & $4.27\times10^{2}$ & $1.00\times10^{4}$ & $2.20\times10^{4}$ \\
$\bm{\mathrm{eZV}_{50,\mathrm{SAvg}}}$ & $4.29\times10^{1}$ & $4.40\times10^{2}$ & $1.07\times10^{4}$ & $2.91\times10^{4}$ & $4.29\times10^{1}$ & $4.38\times10^{2}$ & $9.27\times10^{3}$ & $1.75\times10^{4}$ \\
\midrule
$\bm{\mathrm{eZV}_{25,\mathrm{QP}_{+}}}$ & $4.33\times10^{1}$ & $5.20\times10^{2}$ & $1.05\times10^{4}$ & $3.15\times10^{4}$ & $4.33\times10^{1}$ & $5.18\times10^{2}$ & $9.75\times10^{3}$ & $2.36\times10^{4}$ \\
$\bm{\mathrm{eZV}_{50,\mathrm{QP}_{+}}}$ & $4.18\times10^{1}$ & \textcolor{red}{$\bm{6.48\times10^{2}}$} & \textcolor{red}{$\bm{1.32\times10^{4}}$} & $3.21\times10^{4}$ & $4.18\times10^{1}$ & \textcolor{red}{$\bm{6.44\times10^{2}}$} & \textcolor{red}{$\bm{1.14\times10^{4}}$} & $1.93\times10^{4}$ \\
\bottomrule
\end{tabular}
}
\end{table}

In Experiment 2, as sampling is computationally expensive, the overall efficiency is primarily determined by statistical efficiency, except for the regularised ZVCV methods with higher polynomial orders and the ensemble methods at $S = 3000$. Our proposed approaches achieve the highest statistical and overall efficiency when $S \in \{300, 1000\}$, with performance improving as $k$ increases, but are slightly outperformed by $\text{SECF}_2$ when $S = 100$ and by $\text{ZV}_4$ when $S=3000$, the smallest sample size at which $\text{ZV}_4$ is identifiable. At $S = 1000$, the best ensemble improves on the best non-ensemble method, $\text{ZV}_3$, by a factor of roughly $2.0$, and on the best regularised ZVCV method by a factor of more than 90. We expect their statistical efficiency to improve further with a higher base polynomial order, given that the performance of ZVCV increases with $Q$. The default ensemble size $k=25$ performs similarly to $k=50$, with the largest difference for the QP$_+$-based method at $S \in \{300, 1000\}$, where $k=50$ is about 25\% more efficient. Under the default settings, the QP$_+$-based method slightly outperforms the SAvg-based one at most sample sizes, in contrast to Experiment 1. Coupled with the results from Experiment 1, this suggests that neither combination rule uniformly dominates; the preferable choice appears to depend on the degree of heterogeneity in the variances of the constituent learners.

The regularised ZVCV methods show only modest benefit from increasing the polynomial order beyond $Q = 2$; for example, at $S = 1000$, the statistical efficiency of $r$-$\text{ZV}_Q$ improves from $1.08\times10^{2}$ to $1.42\times10^{2}$ as $Q$ increases from 3 to 5, while its overall efficiency declines from $8.40\times10^{1}$ to $2.71\times10^{1}$. Even when sampling is costly, they remain noticeably slower than the other methods. CF yields the worst results, which are improved upon by SECF. $\text{SECF}$ outperforms ZV at the same polynomial order. Similar to ensemble ZVCV methods, we expect the efficiency of SECF to improve with a higher base polynomial order.

\subsubsection{Experiment 3: Log-Gaussian Cox Process}

\begin{table}[t]
    \renewcommand{\arraystretch}{1}
    \centering
    \caption{Results of Experiment 3: Log-Gaussian Cox process}
    \label{table:cox}
\begin{tabular}{ccccc}
\toprule
& \multicolumn{2}{c}{Statistical efficiency} & \multicolumn{2}{c}{Overall efficiency} \\
\cmidrule(r){2-3}\cmidrule(l){4-5}
Method & $S = 300$ & $S = 1000$ & $S = 300$ & $S = 1000$ \\
\midrule
$\text{ZV}_1$ & 6.01 & 3.66 & 5.56 & 3.57 \\
\midrule
$r\text{-ZV}_1$ & 6.03 & 3.66 & 0.27 & 0.41 \\
$r\text{-ZV}_2$ & 5.51 & 3.47 & 0.01 & 0.00 \\
\midrule
$\ell\text{-ZV}_1$ & 7.07 & 3.74 & 0.35 & 0.47 \\
$\ell\text{-ZV}_2$ & 517.81 & 636.86 & 2.34 & 2.19 \\
\midrule
$\text{CF}$ & 2.35 & 2.77 & 1.23 & 0.65 \\
$\text{SECF}_1$ & 6.47 & 4.26 & 5.66 & 2.62 \\
\midrule
$\text{a-ZV}_{1,4}$ & 5413.09 & 3225.07 & \textcolor{red}{\textbf{4233.06}} & \textcolor{red}{\textbf{2988.13}} \\
$\text{a-ZV}_{k_{\max},4}$ & 8046.00 & 14770.36 & 1061.02 & 198.56 \\
\midrule
$\bm{\mathrm{eZV}_{k_{\max},\mathrm{SAvg}}}$ & 10451.29 & 7428.82 & 977.29 & 60.39 \\
$\bm{\mathrm{eZV}_{k_{\max},\mathrm{QP}_{+}}}$ & \textcolor{red}{\textbf{14599.20}} & \textcolor{red}{\textbf{16859.43}} & 1342.79 & 136.99 \\
\bottomrule
\end{tabular}
\end{table}

Among the methods that use all $d = 64$ dimensions, $\ell$-$\text{ZV}_2$ is the only one that achieves substantial variance reduction, since LASSO can select a small number of relevant covariates from the $J = 2144$ candidates of the second-order basis. Ridge regression does not share this benefit: $r$-$\text{ZV}_2$ is slightly less efficient than $r$-$\text{ZV}_1$, which is consistent with over-shrinkage of the coefficients when $S \ll J$. However, the regularised ZVCV methods have poor overall efficiency, with $\ell$-$\text{ZV}_1$, $r$-$\text{ZV}_1$ and $r$-$\text{ZV}_2$ falling below vanilla Monte Carlo, as does CF at $S = 1000$, because cross-validating a penalised regression with 64 or 2144 covariates is expensive relative to the comparatively fast Cox process sampler. Restricting the Stein basis to a small spatial neighbourhood is far more effective: $\text{a-ZV}_{1,4}$ improves on $\ell$-$\text{ZV}_2$ by a factor of roughly 5 to 10 in statistical efficiency and by more than three orders of magnitude in overall efficiency.

Comparing $\text{a-ZV}_{1,4}$ and $\text{a-ZV}_{m_{\max},4}$, the larger neighbourhood achieves higher statistical efficiency but far lower overall efficiency, since the richer model costs substantially more to fit. The ensemble amplifies this trade-off: $\mathrm{eZV}_{m_{\max},\mathrm{QP}_{+}}$ attains the highest statistical efficiency of any method at both sample sizes, and is roughly 27 times more statistically efficient than $\ell$-$\text{ZV}_2$, yet its overall efficiency trails that of $\text{a-ZV}_{1,4}$, as fitting up to $m_{\max}$ heterogeneous learners per cell is non-negligible relative to the cost of sampling here. The decline in overall efficiency from $S=300$ to $S=1000$ reflects $m_{\max}$ growing from 6 to 9 rather than a genuine degradation. Between combination rules, QP$_+$ outperforms SAvg in statistical efficiency at both sample sizes, consistent with SAvg offering no protection against a poorly performing learner among heterogeneous learners (Remark \ref{rem:weights}). The method $\mathrm{eZV}_{m_{\max},\mathrm{QP}_{+}}$ outperforms both of the single learners it combines in statistical efficiency at every sample size, confirming that combining heterogeneous neighbourhood sizes adds value beyond either extreme alone.

\section{Conclusion}
\label{section:conclusion}

This work has explored the use of averaging-based ensemble learning approaches to construct Stein-based control variates, with a particular focus on ZVCV, which may become over-parameterised in medium- and high-dimensional settings. While regularised ZVCV techniques based on penalised regression techniques help to mitigate this issue, they can be prohibitively expensive. Their performance may depend on the choice of polynomial order. Motivated by the findings of \cite{lejeune2020implicit}, which demonstrate that, under standard regression assumptions, a properly tuned random-forest-style ensemble of OLS estimators can achieve implicit regularisation similar to the optimal ridge regression, we propose (semi-exact) ensemble ZVCV approaches based on averaging-based ensemble learning. Although these assumptions are generally violated in the ZVCV context and our ensemble algorithm differs somewhat, we expect the proposed methods to perform at least as well as regularised ZVCV in terms of statistical efficiency while being computationally faster.

We have conducted a simulation study to compare the proposed approaches with established control variate techniques. The results indicate that ensemble methods generally outperform regularised ZVCV in statistical efficiency while also being substantially faster. In many scenarios, they also outperform ZVCV and kernel-based methods. The default configuration performs well without requiring explicit tuning of the polynomial order, as we include all terms up to a base polynomial order $Q_{\text{base}}$ exactly, and approximate higher-degree terms via sampling. Performance can likely be further improved by increasing $Q_{\text{base}}$ when the sample size $S$ is sufficiently large. In a 64-dimensional example, an ensemble of \textit{a priori} ZVCV learners built from nested spatial neighbourhoods achieves the highest statistical efficiency of all methods considered, although its overall efficiency depends on the cost of sampling relative to fitting.

Our methods are particularly effective in medium-dimensional settings where sampling is expensive, and in higher-dimensional settings when prior information about the relevant dimensions is available. Future work may investigate base ZVCV learners that are more statistically efficient or scalable in higher dimensions. While our study focuses on ZVCV, the ensemble learning framework is broadly applicable and can be extended to other Stein-based control variates, offering a promising direction for future research.

\section{Disclosure Statement}\label{disclosure-statement}

The authors declare that there are no conflicts of interest.

\section{Acknowledgement}

MLN was supported by a QUT Postgraduate Research Award. LFS was supported by an Australian Research Council (ARC) Discovery Early Career Researcher Award (DE240101190). CD was supported by an ARC Future Fellowship (FT210100260). LFS and CD were also supported by the ARC Discovery Project DP260101525. CD was supported by the ARC Centre of Excellence for the Mathematical Analysis of Cellular Systems (CE230100001). The authors gratefully acknowledge the QUT eResearch team for providing access to high-performance computing (HPC) resources and support. We thank the anonymous referees for their helpful feedback on the manuscript.

\section*{Data Availability Statement}

The code and data used in this manuscript is available in the GitHub repository \href{https://github.com/edelweiss611428/ZVengine}{\texttt{\\edelweiss611428/ZVengine}}.
\phantomsection\label{supplementary-material}
\bigskip

\begin{center}

{\large\bf SUPPLEMENTARY MATERIAL}

\end{center}

\begin{description}
\item[Title: R code implementing ensemble ZVCV methods (zip file)]

The zip file contains the code used in the paper, including the implementation of the ensemble ZVCV algorithms, data and Stan code, and scripts for running the simulation study and combining the results.
\end{description}

  \bibliography{bibliography}

\appendix

\section{Pseudocode of QP$_+$}
\label{appendix:combining}

The pseudocode of the ensemble ZVCV algorithm with QP$_+$-based weights is presented in Algorithm \ref{algo:2e}. The LOO residuals use the standard identity $r_{s,i} = e_{s,i}/(1-h_{s,i})$ for OLS, where $e_{s,i}$ and $h_{s,i}$ are the in-sample residual and leverage of the $s$th sample in the $i$th regression, so that no regression is refitted.

\begin{algorithm}[t]
\SetAlgoLined
\DontPrintSemicolon
\caption{The Ensemble ZVCV Algorithm with QP$_+$-Based Weights}
\label{algo:2e}
\KwIn{integrand evaluation $\textbf{f}_S$, learner design matrices $\textbf{Z}_1,\dots,\textbf{Z}_k$ ($\textbf{Z}_i \in \mathbb{R}^{S \times (J^*_i+1)}$, including an intercept column)}
\KwOut{learner estimates $\hat\alpha_1,\dots,\hat\alpha_k$, weights $\mathbf{w}^{\mathrm{opt}}$, ensemble estimate $\hat{I}_{\mathrm{ens}}$}
\For{$i = 1$ \KwTo $k$}{
  fit OLS via QR: $\textbf{Z}_i = \textbf{Q}_i \textbf{R}_i$, $\hat{\bm\beta}_i = \textbf{R}_i^{-1} \textbf{Q}_i^\top \textbf{f}_S$\;
  leverages: $h_{s,i} = \lVert \textbf{Q}_i[s,] \rVert_2^2$, $s = 1,\dots,S$\;
  in-sample residuals: $e_{s,i} = \textbf{f}_S[s] - (\textbf{Z}_i \hat{\bm\beta}_i)[s]$\;
  LOO residuals: $r_{s,i} = e_{s,i} / (1 - h_{s,i})$ \tcp*{no refitting}
  $\hat\alpha_i \gets$ intercept entry of $\hat{\bm\beta}_i$\;
}
$\textbf{R} \gets [r_{s,i}] \in \mathbb{R}^{S \times k}$\;
$\hat{\bm\Sigma} \gets$ sample covariance matrix of the columns of $\textbf{R}$\;
$\mathbf{w}^{\mathrm{opt}} \gets \displaystyle\argmin_{\mathbf{w} \in [0,1]^k} \mathbf{w}^\top \hat{\bm\Sigma} \mathbf{w} \;\text{s.t.}\; \textstyle\sum_{i=1}^k w_i = 1$\;
$\hat{I}_{\mathrm{ens}} \gets \sum_{i=1}^k w^{\mathrm{opt}}_i \hat\alpha_i$\;
\end{algorithm}

\section{Additional Simulation Results}
\label{appendix:additional}

Section \ref{section:sstudy} reports only the stratified variant of the
semi-exact ensembles. Here we compare it against drawing the $J^*-J^{\text{base}}$ higher-order monomials via SRSWOR over $\mathbf{Z} \backslash \mathbf{Z}^{\text{base}}$ (uniform), rather than allocating them
across orders. Both variants are semi-exact: every learner retains all monomials up to order $Q^{\text{base}}$, so Proposition \ref{prop:trivial} applies to
either, and they differ only in how the higher-order monomials are selected. All
other hyper-parameters are fixed at the defaults of Section \ref{section:tuning}.
Results for the Lotka--Volterra and Friberg--Karlsson examples are given in
Tables \ref{table:lv-strat-unif} and \ref{table:fk-strat-unif}, which also include $\mathrm{QP}_{\pm}$ (Section \ref{section:combining}). $\mathrm{QP}_{\pm}$ is outperformed by QP$_+$ at nearly every setting considered: removing the non-negativity constraint enlarges the feasible set but leaves the weights free to grow in magnitude, amplifying the estimation error in the proxy covariance matrix.

Neither selection strategy dominates, and the two perform comparably throughout,
as expected given that they share all base monomials. In the Friberg--Karlsson example stratified selection is preferable at
the smaller sample sizes, performing best on both metrics at $S=300$ and $S=1000$, while
uniform selection overtakes it at $S=3000$; in the Lotka--Volterra example
uniform selection is ahead at $S=1000$. The allocation of the higher-order
monomials therefore matters mainly when they make up a large share of
$J^*$, and its influence diminishes as the sample size grows. At $S=3000$ in
the Lotka--Volterra example the two schemes coincide exactly, since $J^*=1369$
exceeds the full basis size $J=1286$ and every learner selects all available
covariates.

\begin{table}[h]
    \renewcommand{\arraystretch}{1}
    \centering
    \caption{Experiment 1: stratified versus uniform selection of higher-order monomials}
    \label{table:lv-strat-unif}
    \resizebox{0.85\textwidth}{!}{
\begin{tabular}{ccccccccc}
\toprule
& \multicolumn{4}{c}{Statistical efficiency} & \multicolumn{4}{c}{Overall efficiency} \\
\cmidrule(r){2-5}\cmidrule(l){6-9}
Method & $S = 100$ & $S = 300$ & $S = 1000$ & $S = 3000$ & $S = 100$ & $S = 300$ & $S = 1000$ & $S = 3000$ \\
\midrule
$\bm{\mathrm{eZV}^{\mathrm{strat}}_{25,\mathrm{SAvg}}}$ & \textcolor{red}{$\bm{1.08\times10^{2}}$} & $2.03\times10^{3}$ & $5.03\times10^{4}$ & \textcolor{red}{$\bm{4.33\times10^{4}}$} & \textcolor{red}{$\bm{1.08\times10^{2}}$} & $2.00\times10^{3}$ & $3.37\times10^{4}$ & \textcolor{red}{$\bm{3.98\times10^{4}}$} \\
$\bm{\mathrm{eZV}^{\mathrm{strat}}_{25,\mathrm{QP}_{+}}}$ & $8.61\times10^{1}$ & $1.76\times10^{3}$ & $4.76\times10^{4}$ & $4.33\times10^{4}$ & $8.61\times10^{1}$ & $1.73\times10^{3}$ & $3.18\times10^{4}$ & $3.98\times10^{4}$ \\
$\bm{\mathrm{eZV}^{\mathrm{strat}}_{25,\mathrm{QP}_{\pm}}}$ & $7.70\times10^{1}$ & $1.81\times10^{3}$ & $4.44\times10^{4}$ & $4.33\times10^{4}$ & $7.69\times10^{1}$ & $1.78\times10^{3}$ & $2.97\times10^{4}$ & $3.98\times10^{4}$ \\
$\bm{\mathrm{eZV}^{\mathrm{strat}}_{50,\mathrm{SAvg}}}$ & $1.08\times10^{2}$ & \textcolor{red}{$\bm{2.17\times10^{3}}$} & $5.13\times10^{4}$ & $4.33\times10^{4}$ & $1.08\times10^{2}$ & \textcolor{red}{$\bm{2.10\times10^{3}}$} & $2.58\times10^{4}$ & $3.98\times10^{4}$ \\
$\bm{\mathrm{eZV}^{\mathrm{strat}}_{50,\mathrm{QP}_{+}}}$ & $9.59\times10^{1}$ & $2.09\times10^{3}$ & $4.95\times10^{4}$ & $4.33\times10^{4}$ & $9.58\times10^{1}$ & $2.01\times10^{3}$ & $2.49\times10^{4}$ & $3.98\times10^{4}$ \\
$\bm{\mathrm{eZV}^{\mathrm{strat}}_{50,\mathrm{QP}_{\pm}}}$ & $6.86\times10^{1}$ & $1.64\times10^{3}$ & $4.59\times10^{4}$ & $4.33\times10^{4}$ & $6.85\times10^{1}$ & $1.58\times10^{3}$ & $2.31\times10^{4}$ & $3.98\times10^{4}$ \\
\midrule
$\bm{\mathrm{eZV}^{\mathrm{unif}}_{25,\mathrm{SAvg}}}$ & $9.79\times10^{1}$ & $1.78\times10^{3}$ & \textcolor{red}{$\bm{5.74\times10^{4}}$} & $4.33\times10^{4}$ & $9.79\times10^{1}$ & $1.75\times10^{3}$ & \textcolor{red}{$\bm{3.84\times10^{4}}$} & $3.98\times10^{4}$ \\
$\bm{\mathrm{eZV}^{\mathrm{unif}}_{25,\mathrm{QP}_{+}}}$ & $8.84\times10^{1}$ & $1.69\times10^{3}$ & $5.37\times10^{4}$ & $4.33\times10^{4}$ & $8.84\times10^{1}$ & $1.66\times10^{3}$ & $3.60\times10^{4}$ & $3.98\times10^{4}$ \\
$\bm{\mathrm{eZV}^{\mathrm{unif}}_{25,\mathrm{QP}_{\pm}}}$ & $6.59\times10^{1}$ & $1.51\times10^{3}$ & $4.88\times10^{4}$ & $4.33\times10^{4}$ & $6.58\times10^{1}$ & $1.48\times10^{3}$ & $3.27\times10^{4}$ & $3.98\times10^{4}$ \\
$\bm{\mathrm{eZV}^{\mathrm{unif}}_{50,\mathrm{SAvg}}}$ & $1.03\times10^{2}$ & $1.78\times10^{3}$ & $5.73\times10^{4}$ & $4.33\times10^{4}$ & $1.03\times10^{2}$ & $1.72\times10^{3}$ & $2.88\times10^{4}$ & $3.98\times10^{4}$ \\
$\bm{\mathrm{eZV}^{\mathrm{unif}}_{50,\mathrm{QP}_{+}}}$ & $8.95\times10^{1}$ & $1.76\times10^{3}$ & $5.64\times10^{4}$ & $4.33\times10^{4}$ & $8.94\times10^{1}$ & $1.70\times10^{3}$ & $2.84\times10^{4}$ & $3.98\times10^{4}$ \\
$\bm{\mathrm{eZV}^{\mathrm{unif}}_{50,\mathrm{QP}_{\pm}}}$ & $6.23\times10^{1}$ & $1.41\times10^{3}$ & $4.46\times10^{4}$ & $4.33\times10^{4}$ & $6.22\times10^{1}$ & $1.36\times10^{3}$ & $2.25\times10^{4}$ & $3.98\times10^{4}$ \\
\bottomrule
\end{tabular}
}
\end{table}

\begin{table}[h]
    \renewcommand{\arraystretch}{1}
    \centering
    \caption{Experiment 2: stratified versus uniform selection of higher-order monomials}
    \label{table:fk-strat-unif}
    \resizebox{0.85\textwidth}{!}{
\begin{tabular}{ccccccccc}
\toprule
& \multicolumn{4}{c}{Statistical efficiency} & \multicolumn{4}{c}{Overall efficiency} \\
\cmidrule(r){2-5}\cmidrule(l){6-9}
Method & $S = 100$ & $S = 300$ & $S = 1000$ & $S = 3000$ & $S = 100$ & $S = 300$ & $S = 1000$ & $S = 3000$ \\
\midrule
$\bm{\mathrm{eZV}^{\mathrm{strat}}_{25,\mathrm{SAvg}}}$ & $4.28\times10^{1}$ & $4.28\times10^{2}$ & $1.08\times10^{4}$ & $2.94\times10^{4}$ & $4.28\times10^{1}$ & $4.27\times10^{2}$ & $1.00\times10^{4}$ & $2.20\times10^{4}$ \\
$\bm{\mathrm{eZV}^{\mathrm{strat}}_{25,\mathrm{QP}_{+}}}$ & $4.33\times10^{1}$ & $5.20\times10^{2}$ & $1.05\times10^{4}$ & $3.15\times10^{4}$ & $4.33\times10^{1}$ & $5.18\times10^{2}$ & $9.75\times10^{3}$ & $2.36\times10^{4}$ \\
$\bm{\mathrm{eZV}^{\mathrm{strat}}_{25,\mathrm{QP}_{\pm}}}$ & $2.61\times10^{1}$ & $4.66\times10^{2}$ & $1.00\times10^{4}$ & $3.15\times10^{4}$ & $2.61\times10^{1}$ & $4.64\times10^{2}$ & $9.27\times10^{3}$ & $2.37\times10^{4}$ \\
$\bm{\mathrm{eZV}^{\mathrm{strat}}_{50,\mathrm{SAvg}}}$ & $4.29\times10^{1}$ & $4.40\times10^{2}$ & $1.07\times10^{4}$ & $2.91\times10^{4}$ & $4.29\times10^{1}$ & $4.38\times10^{2}$ & $9.27\times10^{3}$ & $1.75\times10^{4}$ \\
$\bm{\mathrm{eZV}^{\mathrm{strat}}_{50,\mathrm{QP}_{+}}}$ & $4.18\times10^{1}$ & \textcolor{red}{$\bm{6.48\times10^{2}}$} & \textcolor{red}{$\bm{1.32\times10^{4}}$} & $3.21\times10^{4}$ & $4.18\times10^{1}$ & \textcolor{red}{$\bm{6.44\times10^{2}}$} & \textcolor{red}{$\bm{1.14\times10^{4}}$} & $1.93\times10^{4}$ \\
$\bm{\mathrm{eZV}^{\mathrm{strat}}_{50,\mathrm{QP}_{\pm}}}$ & $2.56\times10^{1}$ & $6.03\times10^{2}$ & $1.16\times10^{4}$ & $3.21\times10^{4}$ & $2.56\times10^{1}$ & $5.99\times10^{2}$ & $1.00\times10^{4}$ & $1.94\times10^{4}$ \\
\midrule
$\bm{\mathrm{eZV}^{\mathrm{unif}}_{25,\mathrm{SAvg}}}$ & $4.22\times10^{1}$ & $3.75\times10^{2}$ & $1.06\times10^{4}$ & $3.27\times10^{4}$ & $4.22\times10^{1}$ & $3.74\times10^{2}$ & $9.85\times10^{3}$ & $2.47\times10^{4}$ \\
$\bm{\mathrm{eZV}^{\mathrm{unif}}_{25,\mathrm{QP}_{+}}}$ & \textcolor{red}{$\bm{4.36\times10^{1}}$} & $4.01\times10^{2}$ & $1.04\times10^{4}$ & $3.21\times10^{4}$ & \textcolor{red}{$\bm{4.36\times10^{1}}$} & $4.00\times10^{2}$ & $9.63\times10^{3}$ & $2.43\times10^{4}$ \\
$\bm{\mathrm{eZV}^{\mathrm{unif}}_{25,\mathrm{QP}_{\pm}}}$ & $2.64\times10^{1}$ & $3.61\times10^{2}$ & $1.06\times10^{4}$ & $3.39\times10^{4}$ & $2.64\times10^{1}$ & $3.60\times10^{2}$ & $9.85\times10^{3}$ & \textcolor{red}{$\bm{2.55\times10^{4}}$} \\
$\bm{\mathrm{eZV}^{\mathrm{unif}}_{50,\mathrm{SAvg}}}$ & $4.27\times10^{1}$ & $3.84\times10^{2}$ & $1.20\times10^{4}$ & $3.33\times10^{4}$ & $4.27\times10^{1}$ & $3.82\times10^{2}$ & $1.04\times10^{4}$ & $2.02\times10^{4}$ \\
$\bm{\mathrm{eZV}^{\mathrm{unif}}_{50,\mathrm{QP}_{+}}}$ & $4.34\times10^{1}$ & $4.58\times10^{2}$ & $1.15\times10^{4}$ & \textcolor{red}{$\bm{3.53\times10^{4}}$} & $4.34\times10^{1}$ & $4.55\times10^{2}$ & $9.98\times10^{3}$ & $2.14\times10^{4}$ \\
$\bm{\mathrm{eZV}^{\mathrm{unif}}_{50,\mathrm{QP}_{\pm}}}$ & $2.61\times10^{1}$ & $3.80\times10^{2}$ & $1.10\times10^{4}$ & $3.52\times10^{4}$ & $2.61\times10^{1}$ & $3.77\times10^{2}$ & $9.48\times10^{3}$ & $2.14\times10^{4}$ \\
\bottomrule
\end{tabular}
}
\end{table}
\end{document}